\documentstyle[aps,prd,preprint,epsf]{revtex}

\tighten

\catcode`\@=11
\catcode`\@=12

\def\dddot#1{\mathinner{\buildrel\vbox{\kern5pt\hbox{...}}\over{#1}}}

\def\et{{\it et al}}

\def\be{\begin{equation}}
\def\ee{\end{equation}}
\def\bq{\begin{eqnarray}}
\def\eq{\end{eqnarray}}
\def\beq{\begin{eqnarray*}}
\def\eeq{\end{eqnarray*}}

\def\bs{\begin{subequations}}
\def\es{\end{subequations}}
\def\ben{\begin{eqalignno}}
\def\een{\end{eqalignno}}

\def\({\left(}
\def\){\right)}

\begin{document}

\title{Causal perturbation theory in general FRW cosmologies II:
superhorizon behaviour and initial conditions}

\author{G.~Amery${}^{1}$\thanks{
Electronic address: G.Amery$\,$\hbox{\rm @}$\,$damtp.cam.ac.uk},
and E.P.S.~Shellard${}^{1}$\thanks{
Electronic address: E.P.S.Shellard$\,$\hbox{\rm @}$\,$damtp.cam.ac.uk}}

\address{${}^1$ Department of Applied Mathematics and Theoretical Physics\\
Centre for Mathematical Sciences, University of Cambridge\\
Wilberforce Road, Cambridge CB3 0WA, U.K.}

\maketitle

\begin{abstract}
{We describe how causality and energy-momentum conservation constrain
the superhorizon behaviour of perturbation variables in a general
FRW spacetime. The effect of intrinsic curvature upon the horizon
scale is discussed and `white noise' results for the power spectra of 
generic `causally-seeded' superhorizon
perturbations are obtained.  We present a detailed
derivation of the superhorizon energy density power spectrum for
curved universes, and provide elegant mathematical
arguments for a curved universe analogue of the Bessel function
addition theorem.  This yields physical insight into
both the familiar $k^4$ result for flat
space and  the  $K \not= 0$ case: curvature effects are seen to be
evident from as early as half the curvature scale. 
  We consider the implications of these results for the physically
and numerically significant issue of setting consistent initial conditions.}
\end{abstract}
\pacs{PACS number(s): }

\section{Introduction}
\label{initial}

The standard flat universe inflationary plus dark energy 
paradigm 
appears to be remarkable accord with recent CMB experiments
\cite{Jaffe}.  Nevertheless, 
 the confrontation with observation remains
indecisive, not only because of the significant 
experimental uncertainties, but
also because good quantitative accuracy 
has not yet been achieved for competing paradigms such as 
actively sourced  theories, which have so far only 
been extensively studied for cosmic defects in flat FRW ($K = 0$) 
backgrounds \cite{ACSSV,pst,pst2,durrer}. Moreover, even 
for flat universes,  a subsidiary role for defect networks 
complementing the inflationary power spectrum cannot be excluded
\cite{bprs}. The possibilty of searching for distinct defect signatures
in the CMB, such as nonGaussianity \cite{CMBstrings} or in the polarisation
spectra \cite{SZ} may go some way toward resolving these 
remaining uncertainties.  However, in order to have confidence in 
cosmological parameter estimation with a purely inflationary paradigm, 
it will be necessary to constrain the alternative models, including 
both `primordial' and `active' perturbation sources, as well as the 
effects of vector and tensor modes and  $K \not = 0$ 
backgrounds.  Here the combination of intrinsic curvature and defect sources is
particularly interesting,  motivating a study of the superhorizon
behaviour and the  choice of consistent initial conditions for such
perturbations.  Of particular interest is the possibility of a testable 
signature of superhorizon fall-off in `causal' or `active' models in the 
CMB polarisation power spectrum.

The Cauchy problem for cosmological perturbations over a $K = 0$ FRW
 background has been extensively discussed in the literature. 
Using energy-momentum conservation, and assuming that the early
 universe was purely homogeneous
and isotropic, one can show \cite{pa,CS,peeb,vs} that the power spectrum of 
the stresses should be that of white noise ($\propto k^0$), while
that of the density should be suppressed as $k^4$, for wavelengths much 
larger than the horizon; here, $k$ is the comoving wavenumber so this 
the superhorizon condition is $k \ll aH$ where $H={\dot a/a}$ and $a$ is the scale factor.
Assuming that matter undergoes random displacements of length $d_c$ at each
point in space, while satisfying energy-momentum conservation, one can 
estimate that the density power spectrum
 turns over to $k^4$ behaviour at $k_c \sim 2 d_c^{-1}$ \cite{pa}. 
This result may be used \cite{vs,pst} to simplify the
calculations to determine consistent initial conditions for numerical
analyses. One assumes that all perturbations of relevance today come 
from far outside the initial horizon ($k \tau_{\rm i} \ll 1$) so that one may
argue that the strong $k^4$ suppression allows us to set all these 
variables to zero initially. 

For the general $K \not= 0$ FRW cosmologies, these issues remain
unaddressed, although there do exist several analyses of the effect of
curvature on the CMB assuming primordial inflationary perturbations described by
power law ansatzes. Some examples include: the generalisation of the
 Sachs-Wolfe effect to
include curvature effects \cite{peeb}; refs.
 \cite{KamiSper} and \cite{Wilson} treat the
open universe case, respectively using a gauge invariant and a
synchronous gauge formalism --- see also  ref. \cite{as} --- and
ref. \cite{WhitScot} investigated the closed universe case, again
using a gauge invariant approach. 
However, the literature does not contain (to our knowledge) any direct
analysis of the superhorizon effects of energy-momentum conservation
on active and causal seed sources in intrinsically curved FRW spacetimes.  

In this paper we shall therefore describe the manner in which causality and
energy-momentum conservation constrain the superhorizon behaviour of the
perturbation variables in the general FRW  spacetime, in
curvilinear coordinates.  As discussed in our previous paper ---
ref. \cite{Amery3} --- we may describe the
symmetry properties and stress energy conservation laws of these
spacetimes in two equivalent languages corresponding to integral and
differential conservation laws. 
The language of conserved vector
densities $\hat{I}^\mu$ ({\it c.f.} Katz \et \cite{kblb-1}) uses the
conformal geometry of the background spacetime to naturally
 construct integral constraint equations, 
representing strong conservation laws defined over some region of a
spacelike hypersurface.  These reduce to Noether
conservation laws when constructed using the  Killing geometry of
the background. 

Traschen \cite{tras1} obtained such integral constraints using as a background
manifold a maximally symmetric de Sitter spacetime in
quasi--Minkowskian coordinates, and these were related to the vector
densities $\hat{I}^\mu_{\bf \xi}$ by Dereulle \et \cite{dku-1}.   However, in order to make use of
these integral constraints, we need to know their behaviour on the
boundary of the integration volume.  In the case of bounded (or ``local'')
 perturbations, the boundary terms
are assumed to vanish, that is, these previous analyses consider 
 perturbations whose stress energy vanishes beyond the
horizon. This is a stronger assumption than required by the physics,
since only the two-point correlators between these quantities have to
vanish in order to satisfy causality. In other words, a physical
assumption has been made that is perhaps better suited to the analysis of an
explosive event rather than more general causal perturbations, such as 
those seeded by
topological defects, which may persist beyond the horizon ({\it
 e.g.} cosmic strings).  Moreover, the choice of background relative to
which energy-momentum is formally defined  is that of a de
Sitter universe, rather than the FRW background convenient for most
 simulations. 

Hence, we turn to the equivalent
language of energy-momentum pseudo-tensors \cite{Amery3}, outlined in \S \ref{SEC-em}.  Again, the relative
simplicity of the flat space case is complicated by the introduction  of
curvature.  In \S \ref{int-k4sec} we define the two--point correlator,
and discuss causality, while the transformed correlator (with respect to the
 eigenfunctions of the Laplacian) is described in \S
\ref{Tcor-k4sec}.  
The effect of intrinsic curvature upon the horizon scale is dealt with in \S 
\ref{HS-k4sec};  and in \S \ref{wn-k4sec} we obtain results for a
general FRW spacetime analogous to the white noise power spectrum of two
arbitrary functions in flat space. Then, in \S \ref{SEC-SHiso},  we
consider in more detail the superhorizon behaviour of the pseudo-energy.
This not only yields the familiar $k^4$ behaviour 
 for the isotropic ($00$-)mode on superhorizon
scales for $K = 0$, but is also well suited to
 study the $K \not= 0$ case.  We regain the flat space
results in the local limit and discuss the effects of
intrinsic curvature on the isotropic mode in a Helmholtz
decomposition of the density perturbation associated with causal
sources, for both open and closed cosmologies. In \S \ref{INITCON} we
comment on the implications of these results for the choice of
consistent initial conditions. We work in the
 synchronous gauge because of its
ubiquity in numerical simulations and the greater physical
transparency offered by this gauge choice.

\section{Energy-momentum conservation} \label{SEC-em}

We shall consider metric perturbations $h_{\mu\nu}$ about a general 
FRW spacetime 
\begin{eqnarray}
ds^2 = a^2 (\gamma_{\mu\nu}+h_{\mu\nu})dx^\mu dx^\nu\,,
\label{FRWmetric}
\end{eqnarray}
where the comoving background line element in `conformal-polar' coordinates
$(\tau,\,\chi,\,\phi,\,\theta)$ is given by
\begin{eqnarray}
\gamma_{\mu\nu}dx^\mu dx^\nu = - d\tau^2 + \frac{1}{|K|} \left[d\chi^2
+ \sin_K ^2 \chi \left( d\theta^2 + \sin^2\theta\, d\phi^2\right) \right] \; ,
\label{FRWspacemetric}
\end{eqnarray}
with the function $\sin_K \chi$ depending on the spatial curvature
$K$ as 
\begin{eqnarray}
\sin_K\chi =\cases {\sinh \chi\,, & $K<0\,,$\cr
\chi\,, & $K=0\,,$\cr
\sin \chi\,, & $K>0\,.$\cr}  \label{sinK}
\end{eqnarray}
We shall adopt the synchronous gauge defined by the choice 
 $h^{0\mu}=0$, so that 
 the trace is given by $h\equiv h^i_{\;\; i}$ (with the convention throughout
that Greek indices run from 0 to 3 and Latin from 1 to 3).  

Here, $a \equiv a(\tau)$ is the scalefactor, for which we can define
the conformal Hubble factor ${\cal H} =\dot a/a$, with dots denoting
derivatives with respect to conformal time $\tau$.  The evolution of
these background quantities is described by the Friedmann equations 
\begin{eqnarray}
\kappa a^2  \rho = 3 ( H^2 + K),
\hspace{20mm} \kappa a^2 (\rho + 3 p)
= - 6 \dot{H}
\; ,  \label{friedeqns}
\end{eqnarray}
where the background tensor $\bar T_{\mu\nu} = (\rho, p, p, p)$
includes the  dark energy of
the universe (or cosmological constant). 
In an earlier paper, \cite{Amery3}, we have shown that one may use the
Bianchi identities to rewrite the Einstein equations 
$G_{\mu\nu}
\equiv R_{\mu\nu} - \frac{1}{2} g_{\mu\nu} R = \kappa T_{\mu\nu}$
(with $\kappa = 8 \pi G$), to obtain a description of  the 
 perturbed energy-momentum in
terms of a linear tensor 
\begin{eqnarray}
\kappa  \tau^0_{\;\; 0} &=& a^2 \delta  G^0_{\;\; 0} + {\cal H}
\dot{h} - K h \; , \hspace{10mm} 
\kappa  \tau^0_{\;\; i} = \delta G^0_{\;\; i} \; , \nonumber
\\
\kappa  \tau^i_{\;\; j} &=& \delta G^i_{\;\; j} - {\cal H}
\left[ \dot{h}^i_{\;\; j} - \dot{h} \delta^i_{\;\; j} \right] \; . 
\label{tau-def1}
\end{eqnarray}
Note that here we have dropped the deltas in the original definition, 
understanding that the
$\tau^\mu_{\;\; \nu}$ are the perturbed parts of the energy-momentum
pseudo-tensor only.  We have also separated the perturbed  
energy-momentum tensor into two parts: the first order part $\delta T_{\mu\nu}$
incorporates the stress energy of the radiation fluid, baryonic matter,
and cold dark matter, and a  
 contribution $\Theta_{\mu\nu}$ representing the stress tensor
of an evolving defect network or some other causal sources.  This is 
assumed to be small (of order $\delta T_{\mu\nu}$) and `stiff', that is,
its energy and momenta are conserved independently of the rest of the 
matter and radiation in the universe and to lowest order its evolution 
is unaffected by the metric perturbations $h_{\mu\nu}$.

The components (\ref{tau-def1}) may be simply related to
 conserved vector densities $\hat{I}^\mu$ obtained by defining
energy and momenta with respect to a general FRW background
\cite{Amery3}, and therefore represent a locally valid definition of
these quantities.  The conservation equations for these quantities are
\begin{eqnarray}
 \tau^0_{\;\; 0, 0} +  \tau^i_{\;\; 0|i} = 0 \; ,
\hspace{10mm} \tau^0_{\;\; i, 0} +  \tau^j_{\;\; i|j} = 0
\; , \label{tau-cons}
\end{eqnarray}
where the bar denotes covariant differentiation with respect to the
spatial three metric $\gamma_{ij}$. 

The pseudo-energy $\tau_0^0$ is a particularly 
useful variable for tracking the growth of large-structure which is why 
we focus on it in this paper.
 In ref. \cite{Amery3}, we showed that pseudo-energy is proportional to 
the growing mode of the density perturbation on superhorizon scales
whether the universe is dominated by radiation, matter or curvature
(see also \cite{vs,pst}). 
In addition, we briefly discussed the implications of matching
conditions at an instantaneous phase transition, with particular
emphasis on the choice of consistent initial conditions.  However, a
more thorough treatment requires that we consider not the
perturbations themselves, but rather their correlators.

\section{Two-point correlators}

\subsection{Two-point correlators and causality} \label{int-k4sec}

Causality is usually defined in terms of 
 two--point correlation functions, which 
contain all the information pertinent to Gaussian perturbations
\cite{wu-2}.
If two random functions $b_{\mu \nu}$ and $d_{\mu \nu}$
are assumed to be statistically spatially homogeneous and isotropic, then their 
two point correlation function, 
$C_{\mu \nu \sigma \lambda} ({\bf y}, \tau, \tau^\prime) = \langle b_{\mu \nu} ({\bf x}, \tau) \;\; d_{\sigma \lambda} ({\bf x} + {\bf y},
\tau^\prime) \rangle$ 
is a function of time and the `radial' coordinate $|{\bf y}| = \chi_y$ only.  The expansion of
the universe breaks time translation invariance \cite{durrer}, so that we may expect
dependence on both $\tau$ and $\tau^\prime$, rather than just on their
difference $\tau - \tau^\prime$.   As we are constructing a causal theory of
cosmological perturbations, we note that causality imposes the
constraint
\begin{eqnarray}
C_{\mu \nu \sigma \lambda} ({\bf y}, \tau, \tau^\prime) = \langle b_{\mu \nu} ({\bf x}, \tau) \;\; d_{\sigma \lambda} ({\bf x} + {\bf y},
\tau^\prime) \rangle = 0,  \hspace{10mm} {\rm for} \hspace{5mm} \chi_{\bf
y} > \tau + \tau^\prime  \; , \label{def-caus}
\end{eqnarray}  
so that the two-point correlation function has compact support in
${\bf y}$.  
Here we may assume without loss of generality
that $\tau > \tau^\prime$. The fact that the correlator has compact
support in  turn means that the Fourier transform (in flat space) of the
correlator is analytic, as are all the integral transforms with
which we shall be concerning ourselves.  

In particular, we shall be concerned with the power spectra of various
functions, defined as the (Fourier or eigenfunction) coefficient
of the equal time two point correlator of the function with its complex
conjugate: 
\begin{eqnarray}
C_{\mu \nu \mu \nu} ({\bf y}, \tau, \tau) = \langle b_{\mu \nu} ({\bf
  x}, \tau) b_{\mu \nu}^* ({\bf x + y}, \tau) \rangle  = \frac{1}{V} \int {\rm d} V_{\bf x} \;\; b_{\mu \nu} ({\bf
x},\tau) \;\; b^*_{\mu \nu} ({\bf x} + {\bf y}, \tau)
\; ,   \label{def-PS}
\end{eqnarray} 
for an arbitrary function $b_{\mu \nu}$. Implicit in the final 
equality of (\ref{def-PS}) is the
``ergodic hypothesis'' which states that one may interchange spatial
and ensemble averages.  In this case the assumption
of causality may be expressed as
\begin{eqnarray}
C_{\mu \nu \mu \nu} ({\bf y}, \tau, \tau) = 0 , \hspace{10mm} {\rm
for} \hspace{5mm} |{\bf y}| > 2 \tau = 2 \chi_H \; . 
 \label{defn-caus}
\end{eqnarray}
In the remainder of this paper we shall frequently suppress the
(equal) time dependence of the correlators.

\subsection{The Helmholtz decomposed correlator}  \label{Tcor-k4sec}

For perturbations over a curved FRW background, it is usual to employ the Helmholtz decomposition 
using the linearly independent eigenfunctions of the Laplacian
in polar coordinates \cite{as,hu}.
We expand all perturbation quantities in terms of the eigenfunctions
${\bf Q}^{(m)}$, which are the scalar ($m = 0$), vector ($m =
\pm 1$) and tensor ($m = \pm 2$) solutions to the Helmholtz equation
\begin{eqnarray}
\nabla^2 {\bf Q}^{(m)} \equiv \gamma^{ij} {\bf Q}^{(m)}_{|ij} = -k^2
{\bf Q}^{(m)} \; ,  \label{eqn-Helm}
\end{eqnarray}
 where the eigentensor has $|m|$
suppressed indices (equal to the rank of the perturbation). The
explicit representations of these eigenfunctions are given in Appendix
A. They are coveniently labeled in terms of a generalised wavenumber
$q$, and its normalised equivalent $\beta = q / \sqrt{|K|}$,  related
to $k$ via
\begin{eqnarray}
q^2 = k^2 + (|m| + 1) K \; , \hspace{6mm} \Longleftrightarrow
\hspace{6mm} \beta^2 = \tilde{k}^2 + (|m| + 1) \frac{K}{|K|}  \; , 
\label{betak-eqn} 
\end{eqnarray}
where $k = \sqrt{|K|} \tilde{k}$ and, in flat space, is the Fourier wavenumber.

The spectra for flat and open universes ($K \leq 0$) are continuous and
complete for $\beta \geq 0$. For the open universe, this means that
 the normalised scalar eigenvalues are greater
than or equal to unity:
$\tilde{k}^2 \geq 1$. For the closed ($K > 0$) case, the
spectrum is discrete because of the existence of periodic boundary
conditons. For scalar perturbations, we then have $\beta = 3,4,5,...$
since the $\beta = 1,2$ modes are pure gauge \cite{LK}.  These spectra
yield a set of eigenfunctions that is complete,
orthonormal, and sufficient to expand all perturbations produced in a
finite region.

The literature (see, for instance, refs. \cite{hu,LythWosz,GBLLW2}) contains 
several comments
about the open universe bound $\tilde{k}^2 \geq 1$, which could be 
misinterpreted as implying that perturbations can only be generated 
below the curvature scale.  However, the particle horizon can 
exceed the curvature scale (see later) which means that, in principle, 
causal perturbations also can be generated on these large lengthscales. 
We shall now briefly place
these comments in the context of  more general eigenfunction
decompositions.  
If one is attempting to describe the stochastic properties of
inflationary 
cosmologies it may be desirable \cite{LythWosz,GBLLW2} to include
the eigenfunctions associated with $0 \leq \tilde{k}^2 < 1$ and even $\tilde{k}^2 < 0$,
as these may produce effects in the CMB anisotropy patterns --- the
Grischuk-Zel`dovich effect \cite{GBLLW1}.  However, this is only necessary if the
assumption that the inflaton field is in the vacuum breaks down for
some correlation length much larger than the curvature scale and the
effect apparently is not measurable in multi-stage inflation models
\cite{BereHeav}.  Moreover,
these eigenfunctions may be obtained only by analytic continuation and
are not orthonormal, nor are they linearly independent of the $\tilde{k}^2
\geq 1$ modes. For
square integrable functions --- such as those describing the correlators
of causal perturbations --- the eigenfunctions labelled by $\tilde{k}^2 \geq 1$
are sufficient. 

The $\tilde{k}^2 \geq 1$ and $\tilde{k}^2 < 1$ modes
are often referred to as the `subcurvature' and  `supercurvature modes', because
their apparent characteristic wavelength ($\propto 1/\tilde{k}$) is
(respectively) smaller or larger than
the comoving curvature scale. However, this nomenclature is
misleading: the $\tilde{k}^2 \geq 1$ modes
can contain information about 
effects above the curvature scale, in which case they are very 
different to their flat space counterparts.  These subcurvature modes 
are {\it not} therefore confined to subcurvature
scales.  They are complete in that they can be used  to describe 
a perturbation on any lengthscale provided it has 
compact support.  Now since
the horizon can exceed the curvature scale, this completeness of the $\tilde k
>1$ modes ensures their sufficiency in describing the correlations of
causal perturbations on all 
lengthscales and at all times.   

In an FRW universe, the power spectrum (\ref{def-PS}) of any scalar
function of ${\bf y}$ in 
terms of the scalar eigenfunctions $\zeta_{\beta l m} = \Phi^l_{\beta} (\chi_{\bf x}) Y_{lm} 
(\theta_{\bf x}, \phi_{\bf x})$, where we have decomposed the $Q^{(0)}$
using spherical harmonics (see (\ref{met-decomp})).  This amounts to a transform on the 
radial coordinate, and a series expansion on the angular part. In particular we
wish to expand the power spectrum of the 
contracted second derivative of the pseudo-stresses $\tau^{i
\;\;\;\; |j}_{\;\; j|i}$ given in (\ref{tau-def1}) as 
\begin{eqnarray}
 \langle \tau^{i \;\;\;\; |j}_{\;\; j|i} ({\bf x}) \;\; 
 \tau^{p \;\;\;\; |q}_{\;\; q|p} ( {\bf x} + {\bf y}) \rangle = 
\sqrt{\frac{2}{\pi}} \sum _{l,m} \int {\rm d} \mu (\beta) \;\; 
 \Phi^l_\beta (\chi_y) Y_{lm} (\theta_y, \phi_y)  
\langle \tau^{i \;\;\;\; |j}_{\;\; j|i}  \;\; 
 \tau^{p \;\;\;\; |q}_{\;\; q|p} \rangle_{lm} (\beta)  
\; , \nonumber 
\end{eqnarray}
where
\begin{eqnarray}
\langle \tau^{i \;\;\;\; |j}_{\;\; j|i} \;\; 
 \tau^{p \;\;\;\; |q}_{\;\; q|p} \rangle_{lm}
 (\beta) &=& \sqrt{\frac{2}{\pi}} \int {\rm d} V_{\bf y} \;\; \Phi^l_\beta (\chi_y)
 Y^*_{lm} (\theta_y, \phi_y)  \langle \tau^{i \;\;\;\; |j}_{\;\; j|i}
 ({\bf x}) \;\;  
 \tau^{p \;\;\;\; |q}_{\;\; q|p} ( {\bf x} + {\bf y}) \rangle \; . 
\nonumber \\
&=& \sqrt{\frac{2}{\pi}} \int {\rm d} V_{\bf x} \;\; \tau^{i \;\;\;\;
 |j}_{\;\; j|i} ({\bf x}) \;\;  \int {\rm d} V_{\bf y}  \;\; \tau^{p \;\;\;\;
 |q}_{\;\; q|p} ( {\bf x} + {\bf y}) \Phi^l_\beta (\chi_y) Y^*_{lm}
 (\theta_y, \phi_y) \; .  
\nonumber \\
\label{dec-correl}
\end{eqnarray}
Here, in the last expression, we have interchanged integration 
with respect to ${\bf x}$ and ${\bf y}$.
In particular, the isotropic 
$00$-mode is given by
\begin{eqnarray}
\langle \tau^{i \;\;\;\; |j}_{\;\; j|i} ({\bf x}) \;\; 
 \tau^{p \;\;\;\; |q}_{\;\; q|p} ( {\bf x} + {\bf y}) \rangle_{00} (\beta) &=& 
\frac{1}{\sqrt{2 \pi^2}} \int {\rm d} V_{\bf y} \;\; \Phi^0_\beta (\chi_{\bf y}) 
\langle \tau^{i \;\;\;\; |j}_{\;\; j|i} ({\bf x})
 \;\;  \tau^{p \;\;\;\; |q}_{\;\; q|p} ( {\bf x} + {\bf y}) \rangle
 \nonumber \\
&=& \frac{1}{\sqrt{2 \pi^2}} \int {\rm d} V_{\bf x} \;\; \tau^{i \;\;\;\;
 |j}_{\;\; j|i} ({\bf x}) \;\;  \int {\rm d} V_y  \;\; \tau^{p \;\;\;\;
 |q}_{\;\; q|p} ( {\bf x} + {\bf y}) \Phi^0_\beta (\chi_y) \; , 
\nonumber \\
\label{00-dcor1}
\end{eqnarray}
where we have substituted $Y_{00} = 1/\sqrt{4 \pi}$.

\section{The transformed horizon scale and white noise}

\subsection{The transformed horizon scale} \label{HS-k4sec}

In order to develop intuition as well as to investigate the curved
universe superhorizon modes in a Helmholtz decomposition,  we need to establish 
the relationship between $\beta$ and the comoving
horizon scale $\chi_H =\tau$.  This is analogous to the flat space result 
$k \tau \approx 3 \pi/2$ for a Fourier-Bessel mode with a wavelength of 
the order of the horizon. 
 We therefore consider the expansion, 
in terms of eigenfunctions of the Helmholtz equation, of a `top hat' window
function that vanishes for $\chi_{\bf x} > \chi_H$, and is given by 
$A ({\bf x}) = A$ for $\chi_{\bf x} \leq \chi_H$, 
where we choose the arbitrary constant $A = 1/\sqrt8$. 
 The fact that $A ({\bf x})$ is 
defined in terms of the radial coordinate only means that we may 
set $ l = 0$ in the Helmholtz decomposition to obtain
\begin{eqnarray}
A ({\bf x}) = \frac{1}{\sqrt{2 \pi^2}} \int {\rm d} \mu (\beta) \;
\Phi^0_{\beta} (\chi) A_{00} (\beta) \; , \hspace{10mm} 
A_{00} (\beta) = \frac{1}{\sqrt{2 \pi^2}} \int {\rm d} V \; A ( {\bf x}) 
\Phi^0_{\beta} (\chi) 
\; . \label{a00}
\end{eqnarray} 
Now, using ${\rm d} V = \sqrt{- g} \,{\rm d} \chi {\rm d} \Omega$, 
$g = {\rm det} (g_{ij})$, 
and appealing to (\ref{relat-1}), we have
\begin{eqnarray}
A_{00} (\beta) = \left\{
\begin{array}{lll}
\frac{ {\rm sinh}^2 \chi_H}{|K|^{3/2} \sqrt{\beta^2 + 1}} \Phi^1_\beta
(\chi_H) &= \frac{1}{|K|^{3/2} \beta^2 + 1} \left[ \frac{ \sin \beta \chi_H \cosh \chi_H}{\beta} - 
\sinh \chi_H \cos \beta \chi_H \right] \; , & 
K < 0 \; , \\
\frac{\chi^2_H}{\beta} \Phi^1_\beta &= \frac{1}{\beta^2} \left[
\frac{\sin \beta \chi_H}{\beta} - \chi_H \cos \beta \chi_H \right] \; , & 
K = 0 \; , \\
\frac{ {\rm sin}^2 \chi_H}{|K|^{3/2} \sqrt{\beta^2 - 1}} \Phi^1_\beta
(\chi_H) &= \frac{1}{|K|^{3/2} \beta^2 - 1} \left[ \frac{ \sin \beta \chi_H \cos \chi_H}{\beta} - 
\sin \chi_H \cos \beta \chi_H \right]  \; , & 
K > 0 \; ,  
\end{array} \right.
\label{a003}
\end{eqnarray} 
where we have used the explicit expression (\ref{Phi-l1}) for $\Phi^1_\beta$ given in
Appendix A. 

In Fig. \ref{fig-Top2} we give the transformed `top hat'
functions for two values of the horizon $\chi_H$, normalised to
$A_{00}=1$ at $\beta=0$.  When the horizon is
 less than or equal to the comoving curvature scale ($\chi_H \leq 1$),
we see that the open universe transform displays
very similar behaviour to the flat space Bessel transform. 
However, on supercurvature scales ($\chi_H > 1$), the open and flat
universe transforms exhibit different behaviour.  In particular, we shall
choose the first zero of the `top hat' transform 
as a means of describing the relationship
between the horizon scale $\chi_H$ and the corresponding value of the
transformed variable $\beta_H$ (the horizon mode wavenumber).  From these 
we will calculate the value
of $$\alpha \equiv \beta_H \chi_H\,,$$ 
in order to discuss how this relationship is affected
by curvature.  The meaning of this relation is that,
for a given $\chi_H$, those modes
with $\beta \chi_H < \alpha$ will have characteristic wavelengths
larger than the horizon. Of course, for flat space and Cartesian coordinates, this
relation is $\lambda = 2 \pi / k < 2 \pi / k_H = \chi_H$. 

\begin{figure}
\leavevmode\epsfxsize=7.5cm \epsfbox{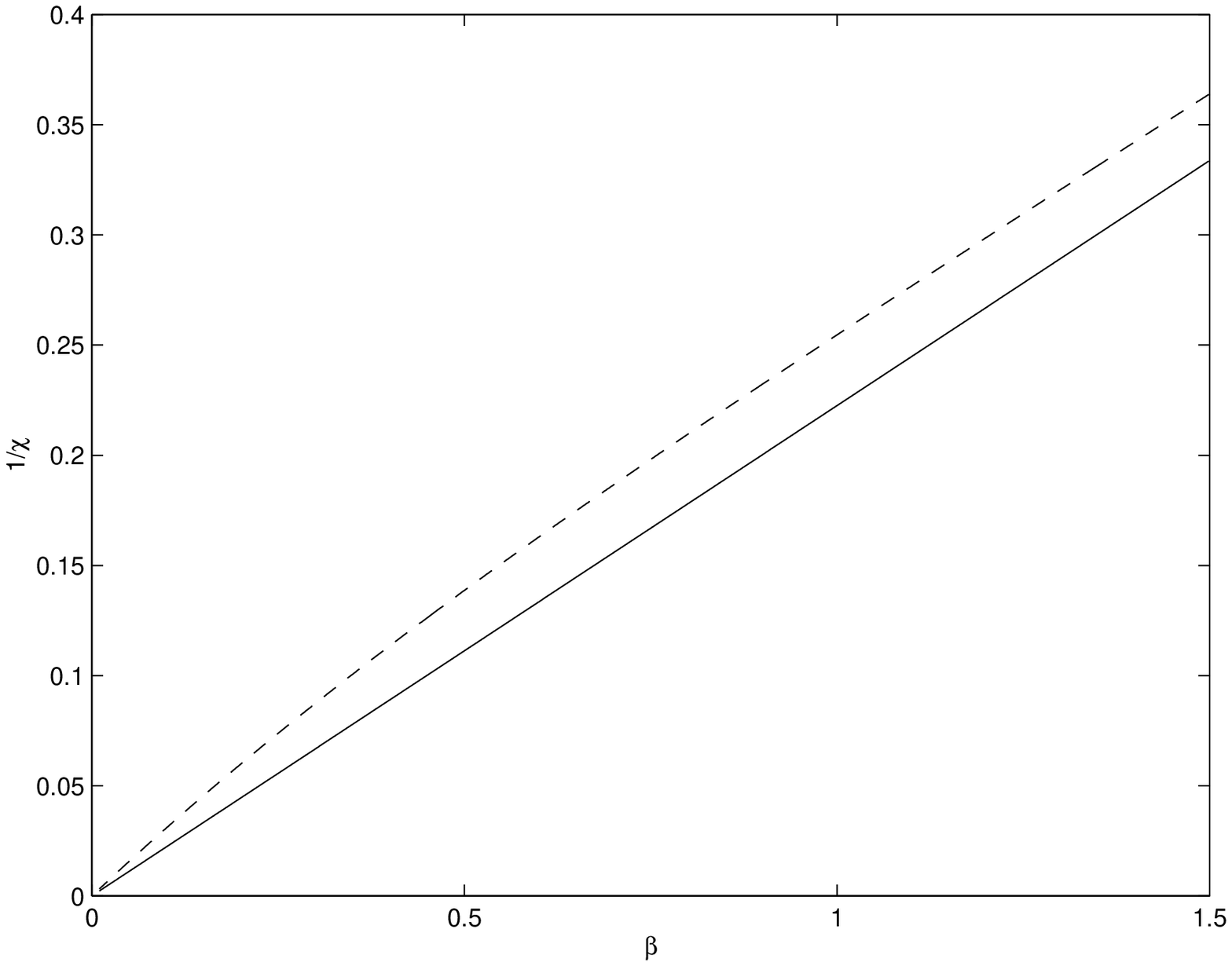}
\leavevmode\epsfxsize=7.5cm \epsfbox{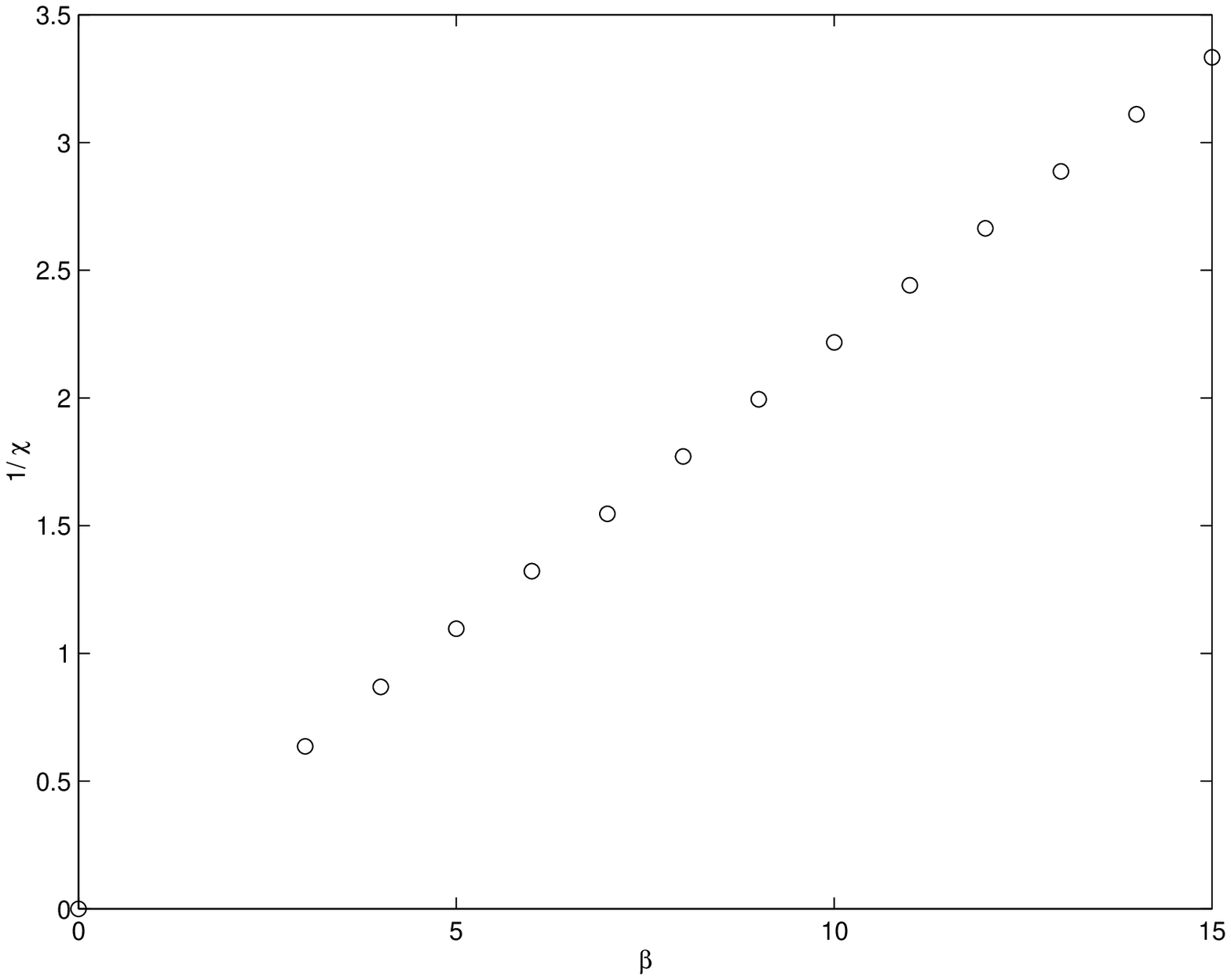}\\ 
  \caption[]
  {Solutions to (\ref{a005}):  $K = 0$ (--), $K = -1$ (- -), $ K = +1$ (o) .}
  \label{fig-f0-1}
\end{figure}

The first zeros of  $A_{00}$  in (\ref{a003}) are given 
by the solutions to 
\begin{eqnarray}
0 = \tan \beta \chi_H - \beta \times \left\{
\begin{array}{ll}
\tanh \chi_H \; , & K < 0 \; , \\
\chi_H \; , & K = 0 \; , \\
\tan \chi_H \; , & K > 0 \; , 
\end{array} \right.
\label{a005}
\end{eqnarray}
which are shown in Fig. \ref{fig-f0-1}.
Note from (\ref{a003}) that the zeros of (\ref{a005}) will be precisely
the zeros of $\Phi^1_\beta$, assuming  $\chi_H > 0$ (for $K \leq 0$) and assuming $0 <
\chi_H < \pi$ for the closed ($K > 0$) case.

The open and flat universes display
phenomenologically similar curves.  Certainly, as can
be seen from Fig. \ref{fig-f0-1}, the curves are approximately parallel
for $\chi_H$ less than (or of the order of) the curvature scale, and
that, on these scales, $\alpha$  is therefore the same ($\sim 3 \pi / 2$) 
as for flat space:  in the local limit, $\beta \chi = q y \longrightarrow k r$ 
 when $\chi \ll 1$ ($y \ll \sqrt{|K|}$) or when $|K|
\longrightarrow 0$.   However, for comoving distances much larger than $\chi_H
\sim 4$ the effect of the curvature is to effectively shrink
 $\alpha$ as the curvature induced volume growth becomes significant. 
The corresponding limit is 
$$
\alpha \equiv \beta_H\chi_H \longrightarrow \pi \qquad \hbox{as}
\qquad \chi_H\longrightarrow \infty \qquad (K<0)\,.
$$  
It is clear from this simple discussion that the characteristic lengthscale
of a mode corresponding to the eigenvalue $k$ is given by 
$1/\beta$.

\begin{figure}
\leavevmode\epsfxsize=7.5cm \epsfbox{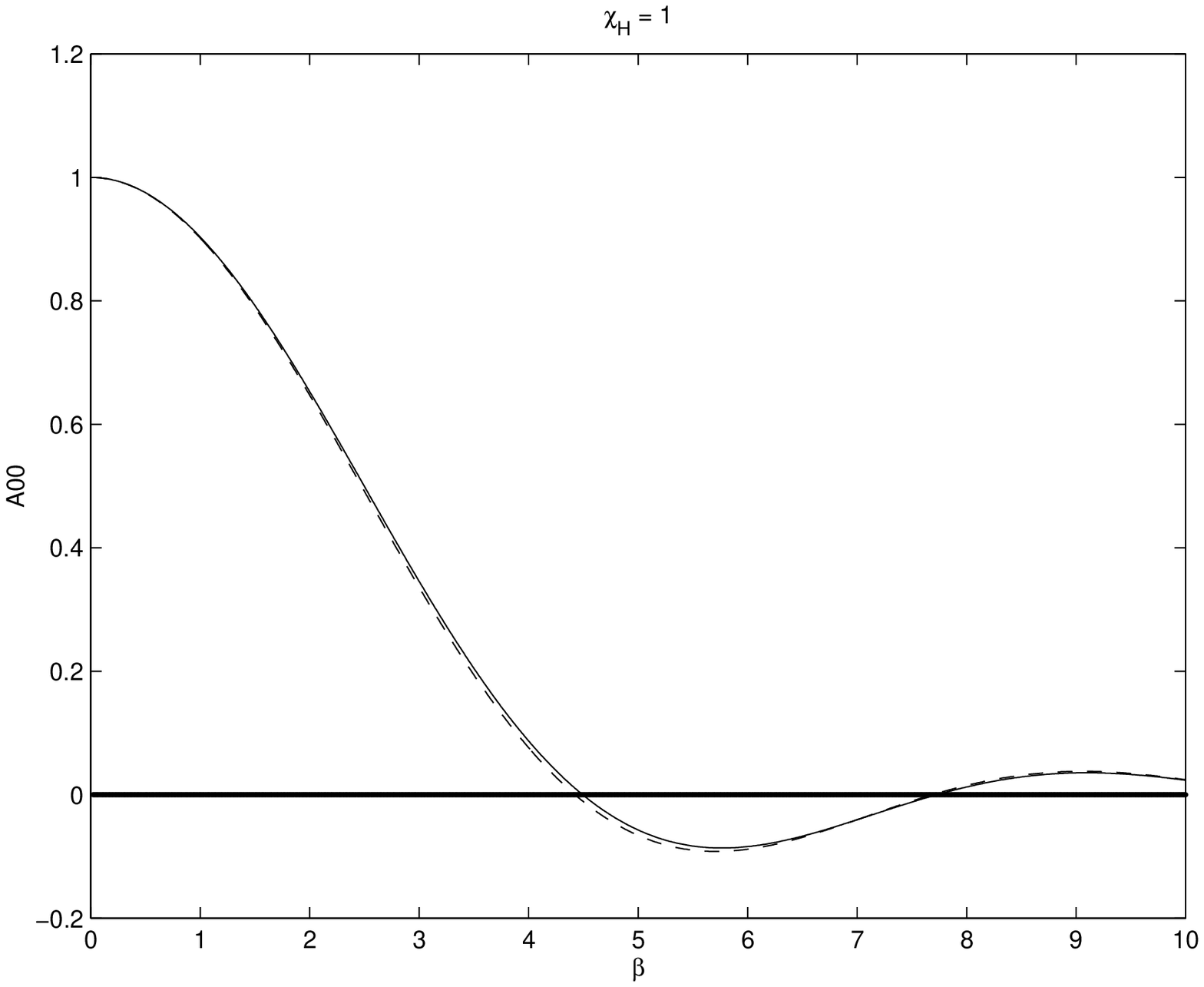}
\leavevmode\epsfxsize=7.5cm \epsfbox{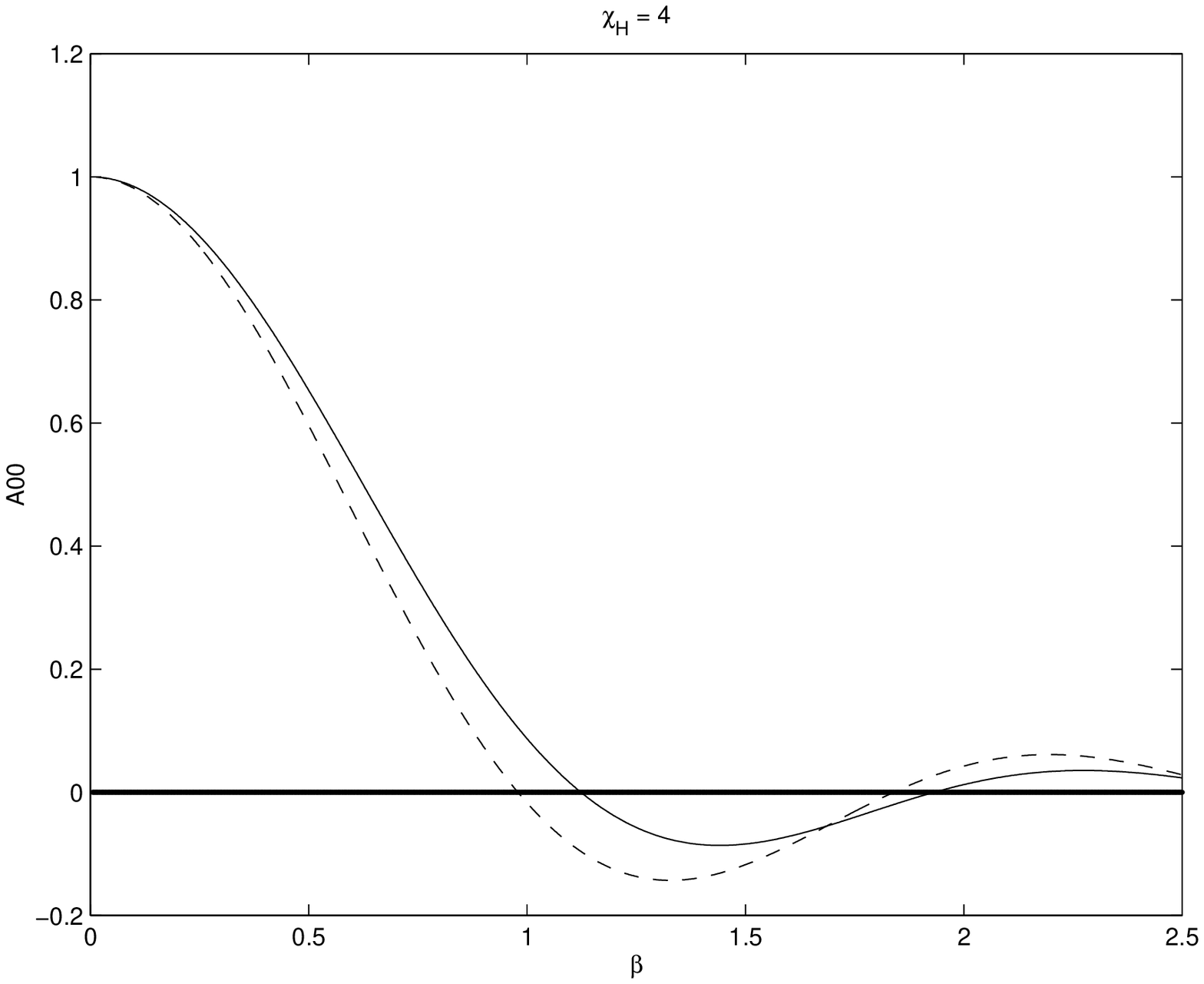}\\ 
  \caption[]
  {The transformed `top hat' function A00: $K = 0$ (--), $K = -1$ (- -).}
  \label{fig-Top2}
\end{figure}

For an open universe, on very subhorizon or very
superhorizon scales, we may simply take $\beta \tau \gg 1$, or $\beta
\tau \ll 1$ respectively, which is directly analogous to the flat space
($\tilde{k}\tau \gg 1$, $\tilde{k}\tau \ll 1$) limits.  However, we note 
that we must not neglect 
the precise location of the horizon if we are to describe the effect of the
curvature scale upon the behaviour of the eigenfunctions. 
We shall
return to this issue later (see \S \ref{SEC-SHiso}), when we observe that 
curvature effects are significant from as
early as $\tau = \chi_H = 0.5$.  

At this point a brief comment on the difference between the
Hubble radius and the particle horizon is relevant. 
In practice, causal perturbations are usually suppressed beyond 
the Hubble radius because this is roughly the maximum distance 
over which correlations
can be established in one Hubble time.  However, in principle, they can 
exist out to the particle horizon, so we must consider this distinction 
carefully (e.g. for cosmic strings or, in a
more extreme case, for exploding neutrino shells).
 From the Friedmann
 equations
(\ref{friedeqns}) we see that the comoving Hubble radius ${\cal H}^{-1} = ( 1 -
\Omega)^{1/2} < 1$ for an open universe, where the density parameter 
$\Omega = \rho / \rho_{\rm crit}$ and
$\rho_{\rm crit} = 3 {\cal H}^2 / \kappa a^2$.   Clearly, this implies that 
the Hubble radius is
always less than the curvature scale.  In inflationary scenarios the
Hubble radius is usually the most important quantity and this bound
has been used to argue that perturbations with
characteristic wavelengths smaller than the curvature scale are well
modelled by a flat universe (see, for example, ref. \cite{LythWosz}).  
However, the open universe particle horizon
is obtained from the equation 
\begin{eqnarray}
\sinh^2 \left( \frac{\chi_H}{2} \right) = \frac{1}{\Omega} - 1 \; ,
\label{parthor2}
\end{eqnarray}
so that the proper distance $d_H = a \chi_H \approx 2 H^{-1}$ for $\Omega
\approx 1$, where $H^{-1}$
is the physical Hubble radius.  It is therefore possible for the particle
horizon to cross the curvature scale: $\chi_H > 1$ for
$\Omega < 0.75$.  Moreover, in order for $\chi_H > 0.5$ today (when
curvature effects become more important --- see \S \ref{SEC-SHiso})  
we require only that $\Omega <
0.94$ today.  This is well within current observational constraints, and
serves to motivate a more complete treatment of the physics around the
 curvature scale.

 For the closed ($K > 0$) case we see that for the discrete 
spectrum $\beta = 3, 4, ... $,
we have  $\alpha \approx 3 \pi/2$, which goes over to the (continuous)
 flat space result in the limit $\chi \rightarrow 0$. 
We note that the notion of supercurvature and superhorizon perturbations with
 $\beta \ll \alpha$, $\chi_H \gg 1$ does not have meaning in the closed universe.
 Indeed, from Fig. \ref{fig-f0-1}, we observe that all the $\beta$ 
 have zeros for values of $\chi_H \leq \pi /2$, at which point the
 closed universe has its maximal circumference and surface area.  

\subsection{The white noise spectrum} \label{wn-k4sec}

We may  use causality and the superhorizon limit to establish that
the correlators between pseudo-stress must go as a constant on
superhorizon scales $\beta \chi_{\bf y} \ll \alpha$.
Substituting the small argument series expansion of $\sin \beta \chi$,
in the exact expression (\ref{Phi-l0}) for 
$\Phi^0_\beta (\chi)$, 
 we find that the isotropic ($l = 0$) mode for the
correlator of two arbitrary scalar functions may be written as  
\begin{eqnarray}
\langle f ({\bf x})  \;\; g ( {\bf x} + {\bf
y}) \rangle_{00} (\beta) = \sum^\infty_{n = 0}
\frac{(-1)^n \beta^{2n}}{(2 n + 1)!}  \int {\rm d} V_{\bf y} \;\; < f
({\bf x})  \;\; g ( {\bf x} + {\bf y}) > 
\times \left\{
\begin{array}{ll}
 \frac{\chi_{\bf y}^{2n + 1}}{
{\rm sinh} ( \chi_{\bf y})} & K < 0 \\
 \chi_{\bf y}^{2n} & K = 0 \\ 
 \frac{\chi_{\bf y}^{2n + 1}}{
{\rm sin} ( \chi_{\bf y})} & K > 0  
\end{array} \right. \; .  
\label{a00b002}
\end{eqnarray}

On superhorizon scales,  we have  $\beta \ll \alpha / \chi_{\bf y}$ for a
given comoving radius $\chi_{\bf y} \gg \chi_{H}$ (where
the results of \S \ref{HS-k4sec} imply that $ \pi <  \alpha
< 3 \pi /2$). On
sufficiently superhorizon scales,
we may therefore choose $\beta \ll 1$, and the lowest order contribution 
 to (\ref{a00b002}) will be proportional to $\beta^0$. Hence, 
$\langle  f ({\bf x}) \;\; g ( {\bf x} + {\bf
y}) \rangle_{00} (\beta) \propto \beta^0$, provided that the integrals
are finite, which will be the
case if we appeal to causality to set $< f ({\bf x})
\;\;  g ( {\bf x} + {\bf
y}) > = 0$ for large values of the radial variable $\chi_{\bf y}$. This
result confirms those by Avelino \cite{pa}, Pen \et \cite{pst}  and
Veerarghavan and Stebbins \cite{vs} for $K = 0$.  

Moreover, as the vector and tensor eigenfunctions given in Appendix A
are proportional to the scalar radial functions $\Phi^l_\beta$, we can
conclude that correlators of vectors and tensors will be premultiplied
by a factor behaving as for the scalars.  This
is in agreement with  the results of Durrer
\et \cite{durrer} for $K
= 0$, and corrects the common, but erroneous, conclusion that the flat
space anisotropic
stress power spectra go as $k^4$ (see, for example, 
refs.~\cite{dku-1,MAFC,hsm}). 

Note that, if we attempt to find the behaviour with respect to $k$, we run into
 difficulties in describing the higher order corrections.  This is a
consequence of the fact that there will be a constant (curvature) 
contribution to the $k^0$ term  from each term $\beta^{2n}$ in the
series expansion (\ref{a00b002}) of $\Phi^0_\beta (\chi_{\bf y})$.

\section{Superhorizon behaviour of the isotropic mode} \label{SEC-SHiso}

Of the several arguments in the literature which use energy-momentum 
 conservation to yield a $k^4$ superhorizon 
power spectrum for flat universes,  that provided by
 Avelino \cite{pa} is the most convincing --- see also refs. 
 \cite{pst,peeb,vs}.  Using Cartesian coordinates, 
 and assuming that the volume is large enough to justify the ergodic
 hypothesis, we may equate spatial and ensemble averages, and Fourier
 transform the correlator of the stresses, arriving at
\begin{eqnarray} 
| \tau_{ij} (k, \eta) \;\; \tau_{kl}^* (k, \eta^\prime)| = V^{-1}
\int_{|{\bf y}| < d_c} {\rm d}^3 y \;\; e^{\imath {\bf k} \cdot {\bf y}}
\langle \tau_{ij} ({\bf x}, \eta) \;\; \tau_{kl} ({\bf x} + {\bf y},
\eta^\prime) \rangle \;.  
\end{eqnarray}
Here we have implicitly made use of the
 convolution theorem, and hence the property $e^{\imath {\bf k} \cdot
 {\bf y}} = e^{- \imath {\bf k} \cdot {\bf x}} e^{\imath {\bf k} \cdot
 ({\bf x + y})}$ of the exponential kernel used in Fourier theory, while
  causality implies that the correlator has
compact support in ${\bf y}$. Hence, on superhorizon scales ($|
 {\bf k} | d_c \ll 1$)  we may neglect the exponential part of the
 integrand and find that the
``power spectrum'' of the pseudo--stresses ${\cal P}_{\tau_{ij}} = V
|\tau_{ij}|^2 / (2 \pi)^3$ approachs a constant: we have a white noise
($k^0$) spectrum.  
Noting that partial differentiation with respect to $x^i$ is
equivalent to multiplying the associated Fourier variable by $k_i$, we
may use energy-momentum conservation (\ref{ped-4}) to obtain the result:
\begin{eqnarray}
{\cal P}_{\tau_{00}} = V \frac{|\tau_{00} (k, \eta)|^2}{(2 \pi)^3} &=&
\frac{k_i k_j k_k k_l}{(2 \pi)^3} \int_{\eta_i}^\eta {\rm d} \eta_1
\int_{\eta_i}^{\eta_1} {\rm d} \eta_2 \int_{\eta_i}^{\eta} {\rm d}
 \eta_1^\prime \int_{\eta_i}^{\eta_1^\prime} {\rm d} \eta_2^\prime
 \times \nonumber \\
& & \int_{|{\bf y}| < d_c} {\rm d}^3 y \;\; \langle \tau_{ij} ({\bf x},
 \eta_2)  \;\; \tau_{kl} ({\bf x} + {\bf y}, \eta_2^\prime) \rangle
\; .  \label{ped-4}
\end{eqnarray}
We conclude that, since $\langle\tau_{ij}\tau_{kl}\rangle$ exhibits 
white noise, then the power spectrum of $\tau_{00}$ should
approach a $k^4$ spectrum on superhorizon scales, as has been
confirmed in numerical studies \cite{RobWan}. Similarly, the power
spectrum of $\tau^0_{\;\; i}$ is proportional to $k^2$, in this
limit. 

The difficulty in generalising the above analysis to the general
FRW universe lies in the lack of a useful
equivalent to the exponential partial wave function 
in the general curved ($K \not= 0$) FRW
universe (see ref. \cite{GBLLW1}), as well as in this argument's 
 dependence upon the choice of a Cartesian coordinate system.
However, we can recast this argument using the eigenmodes of the
Laplacian to investigate the constraints imposed by
energy-momentum conservation and causality in both flat and
instrinsically curved  FRW
universes, regaining also the familiar $k^4$
spectrum in the local (or $|K| \longrightarrow 0$) limit. 

\subsection{Eigenmode splitting in curved spherical coordinates}

In order to generalise this argument we must be able to split our 
eigenmodes in spherical coordinates under spatial translations to 
obtain an analogue of the convolution theorem.
For the flat ($K = 0$) FRW universe  and spherical coordinates we have
the relationship
\begin{eqnarray} 
j_0 (k r_{\bf y}) = \sum_{n=0}^{\infty} (2n + 1) j_n (k r_{\bf x}) 
j_n (k r_{{\bf x}+ {\bf y}}) P_n (\cos \sigma) \; , \label{j0-genbreak}
\end{eqnarray}
where $\sigma$ is the angle between $- {\bf x}$ and ${\bf x} + {\bf
y}$ and we have used $r_{\bf -x} = r_{\bf x}$ \cite{abram}.  In
Appendix B we generalise this to the isotropic radial modes
$\Phi^0_\beta (\chi_{\bf y})$ associated with general FRW spacetimes:
\begin{eqnarray}
\Phi^0_\beta (\chi_{\bf y}) = \sum^\infty_{n = 0} (2 n + 1)
\Phi^n_\beta (\chi_{\bf x}) \Phi^n_\beta (\chi_{{\bf x}+{\bf y}}) P_n (\cos
\sigma) \; . 
\label{Phi0-genbreak}
\end{eqnarray}

The result (\ref{Phi0-genbreak}) is valid for all ${\bf y}$. We are,
however, principally interested in the superhorizon limit, under which
this relation simplifies greatly. In the superhorizon
limit, it is easy to show that  $\Phi^l_\beta (\chi) \propto (\beta \chi)^l$.
Also, ${\bf y}$ superhorizon ($\beta \chi_{\bf y} \ll 1$) implies that
at least one of $- {\bf x}$, ${\bf x}
+ {\bf y}$ is superhorizon also, for any given ${\bf x}$. Hence we
can 
make use of the first order approximation only:
\begin{eqnarray}
\Phi^0_\beta (\chi_{\bf y}) \approx \Phi^0_\beta (\chi_{\bf x})
 \Phi^0_\beta (\chi_{{\bf x} + {\bf y}}) \; ,
\label{phi-split}
\end{eqnarray}
for ${\bf y}$ superhorizon.  
Thus the lowest radial eigen-mode plays a role
analogous to that of the exponential in Fourier theory.  

While Appendix B provides a formal proof of the full expansion 
(\ref{Phi0-genbreak}), here we just obtain (\ref{phi-split}).  
Consider 
an eigenfunction expansion of $\Phi^0_{\beta} (\chi_{\bf u + v})$ in terms of 
$\Phi^0_{\beta} (\chi_{\bf u})$,  after translation to ${\bf
u} + {\bf v}$,  which is given by 
\begin{eqnarray}
\Phi^0_{\beta} (\chi_{{\bf u} + {\bf v}}) &=& \sum_{l,m}
\;\; a^{\beta}_{lm} ({\bf u}) \Phi^l_{\beta} (\chi_{\bf v}) Y_{lm}
(\theta_{\bf v}, \phi_{\bf v}) \; , \nonumber \\
a^{\beta}_{lm} ({\bf u}) &=& \frac{1}{N_l} \int {\rm d} V_{\bf v} \;\;
\Phi^0_{\beta} (\chi_{{\bf u}
+ {\bf v}}) \Phi^l_{\beta} (\chi_{\bf v}) Y_{lm} (\theta_{\bf v}, \phi_{\bf
v}) \; , \nonumber 
\end{eqnarray}
where $N_l = \sqrt{4 \pi} \int {\rm d}
V_{\bf v} \; \left[ \Phi^l_{\beta}
(\chi_{\bf v}) Y_{lm} (\theta_{\bf v}, \phi_{\bf v}) \right]^2$.  We shall demonstrate
that $a^\beta_{00} (\chi_{\bf u}) = \sqrt{4 \pi} \Phi^0_{\beta} (\chi_{\bf u})$,
 so that (for ${\bf u}=-{\bf x}$ and ${\bf v} = {\bf x}+{\bf y}$) eqn~(\ref{phi-split}) is confirmed, on superhorizon scales where
this coefficient carries the leading order behaviour.  Here, we give 
 only the open universe argument, that for the $K \geq
0$ cosmologies being similar.  Substituting for the volume
element, for the $l = 0$ spherical harmonic, and for each of the two
radial eigenfunctions using (\ref{Phi-l0}), yields
\begin{eqnarray}
a^{\beta}_{00} (\chi_{{\bf u} + {\bf v}}) = \frac{\sqrt{\pi}}{|K|^{3/2} \beta^2 N_0}
 \int^\infty_0 {\rm d} \chi_{\bf v} \;\; {\rm sinh} \chi_{\bf v} {\rm
sin} (\beta \chi_{\bf v})   \int^\pi_0 {\rm d} \theta \;\; {\rm sin} \theta
\frac{{\rm sin} (\beta \chi_{{\bf u} + {\bf v}})}{ {\rm sinh} \chi_{{\bf
 u}+{\bf v}}} \; . \nonumber 
\end{eqnarray} 
Changing variables
from $\theta$ to $\xi \equiv \chi_{{\bf u} + {\bf v}}$ using (\ref{eep-1.1}),
we obtain the relation ${\rm sin} \theta {\rm d} \theta =
{\rm sinh} \xi {\rm d} \xi / {\rm sinh} \chi_{\bf v} {\rm sinh}
\chi_{{\bf u}}$, and  calculate the inner integral to be 
 $\frac{2 {\rm sin} (\beta \chi_{\bf v}) {\rm sin} (\beta \chi_{{\bf
u} + {\bf v}})}{\beta {\rm sinh} \chi_{\bf v} {\rm sinh} \chi_{\bf u}
}$. Hence, 
\begin{eqnarray}
a^\beta_{00} (\chi_{\bf v}) 
&=& \frac{2 \sqrt{\pi}}{|K|^{3/2} \beta^2 N_0} \frac{ {\rm sin} (\beta
\chi_{{\bf u}}) }{\beta {\rm sinh} \chi_{{\bf u}}
} \int^\infty_0 {\rm d} \chi_{\bf v} \;\; {\rm sin}^2 \beta \chi_{\bf v}
= \sqrt{4 \pi} \frac{ {\rm sin} \beta \chi_{{\bf u}}}{\beta {\rm sinh}
\chi_{{\bf u}}} \nonumber \\
&=& \sqrt{4 \pi} \Phi^0_\beta (\chi_{{\bf u}}) \; ,
 \label{a00-2}
\end{eqnarray}
where we have used the normalisation $N_0 = ({|K|^{3/2}
\beta^2})^{-1} \int^\infty_0  {\rm d} \chi \; {\rm sin}^2 \beta \chi$ in the
last equality.  Observe that, formally, we should evaluate $N_0$, and effect the
cancellation in (\ref{a00-2}), in terms
of a limit of a volume larger than all length scales of interest. 

\subsection{Integration by parts}

Substituting (\ref{phi-split}) in (\ref{00-dcor1}), transforming the inner integration
variable to ${\bf x} + {\bf y}$, and noting that the
integrals are over all space so that  ${\rm d} V_{{\bf x} + {\bf y}} =
{\rm d} V_{\bf y}$, we see that the $l = 0$ mode is given by the
double integral in ${\bf x}$ and ${\bf x} + {\bf y}$:
\begin{eqnarray}
\langle \tau^{i \;\;\;\; |j}_{\;\; j|i} ({\bf x}) \;\; 
 \tau^{p \;\;\;\; |q}_{\;\; q|p} ( {\bf x} + {\bf y}) \rangle_{00} (k) &=& 
\frac{1}{\sqrt{2 \pi^2}} \int {\rm d} V_{\bf x}  \;\; \tau^{i \;\;\;\; |j}_{\;\; j|i} ({\bf x}) 
\Phi^0_\beta ( \chi_{\bf x}) \nonumber \\
& & \times \int {\rm d} V_{{\bf x} + {\bf y}} \;\;  \tau^{i \;\;\;\; |j}_{\;\; j|i} 
( {\bf x} + {\bf y}) \Phi^0_\beta (\chi_{{\bf x} + {\bf y}}) \; .  
\label{00-3-correl}
\end{eqnarray}
In order to evaluate the integrals in (\ref{00-3-correl}) we make use
of a generalised form of Gauss's Law \cite{stephani}
\begin{eqnarray}
\int_{G^3} W_a^{\;\; ;a} {\rm d} V = \int_{G^2} W_a {\rm d S^a} \; , 
\label{gauss-1}
\end{eqnarray}
where ${\rm d} V = \sqrt{-g^{(3)}}\, {\rm d} x^1   {\rm d} x^2 {\rm d} x^3$ and 
${\rm d} S^a = g^{(3)  ab} {\rm d} S_a$, ${\rm d} S_a = (1/2!) \epsilon_{a m_1 m_2} 
{\rm d}_1 x^{m_1} {\rm d}_2 x^{m_2}$.  Here, we may choose the $3$-dimensional vector 
components  ${\rm d}_i x^{m_i}$ such that ${\rm d} x^{m_i}$ appears as the $i$-th component 
while the rest vanish (for instance, ${\rm d}_2 x^{m_2} = (0, {\rm d} x^{m_2}, 0)$). As we 
are on a constant time hypersurface, the index $a$ runs over $1, 2, 3$ only, 
as do the $m_i$. 

Now, if we replace $W^a$ with $f W^a$ in (\ref{gauss-1}) where $f$ is some scalar quantity
 (so that $f_{;a} = f_{,a}$) and apply the chain rule, we obtain the result
\begin{eqnarray}
\int_{G^3} {\rm d} V f W_a^{\;\; ;a} &=& \int_{G^3} {\rm d} V (f W_a)^{;a} - \int_{G^3} 
{\rm d} V f^{,a} W_a \nonumber \\
&=& \int_{G^2} {\rm d} S^a (f W_a)   - \int_{G^3} {\rm d} V f^{,a} W_a
\; . \label{ibp-1}
\end{eqnarray}
This is the formal expression of spatial integration by parts on a constant time 
hypersurface, and a similar result holds with the covariant and
contravariant indices interchanged.  We shall exploit this result
twice in each of the two integrals in (\ref{00-3-correl}).  

We shall proceed in some detail for the integral over ${\bf x}$. The
${\bf x} + {\bf y}$ integral is similar. Since the $\tau^i_{\;\; j}$ transform as tensors,
setting $f = \Phi^0_\beta (\chi_{\bf x})$ and $W_j =
\tau^i_{\;\; j|i}$ so that $f W_j$ is a vector, we find that  
\begin{eqnarray}
\int {\rm d} V_{\bf x} \;\; \Phi^0_\beta (\chi_{\bf x}) \tau^{i \;\;\;\; |j}_{\;\;
j|i} &=& \int {\rm d} 
V_{\bf x} \Phi^0_\beta (\chi_{\bf x}) W_j^{\;\; |j} \nonumber \\
&=& \int {\rm d} V_{\bf x} \;\; \left[ \Phi^0_\beta (\chi_{\bf x}) W_j \right]^{|j} -
\int {\rm d} V_{\bf x} \;\; \left[ \Phi^0_\beta (\chi_{\bf x}) \right]^{,j} W_j
\nonumber \\
&=&
\int {\rm d} f_{\bf x}^j \;\; \left[ \Phi^0_\beta (\chi_{\bf x}) W_j \right] - 
\int {\rm d} V_{\bf x} \;\; \left[ \Phi^0_\beta (\chi_{\bf x}) \right]^{,j} W_j
\nonumber \\
&=& - \int {\rm d} V_{\bf x} \left[ \Phi^0_\beta (\chi_{\bf x}) \right]^{,j}
\tau^i_{\;\; j|i} \; , 
\label{integral-1}
\end{eqnarray}
where the third equality is obtained as in (\ref{ibp-1}), and 
the fourth equality is obtained by substituting for $W_j$, and assuming that the surface
integral term vanishes.  Explicitly, the surface integrals vanish for
all but the angular integral over the radial surface at infinity, by symmetry,
and periodicity.  This remaining surface integral  may be also be set to zero
on physical grounds by using the conservation equations
(\ref{tau-cons}) to obtain
\begin{eqnarray}
\lim_{R \longrightarrow \infty} \Phi^0_\beta (R) \int {\rm d} \Omega
\left[  \tau^i_{\;\; j|i} \right]_R = \lim_{R \longrightarrow \infty}
\Phi^0_\beta (R) \int {\rm d} \Omega \left[ \dot\tau^0_{\;\; j} \right]_R \; , 
\nonumber 
\end{eqnarray}
and by noting that beyond the horizon the functions $[\Phi^0_\beta \dot{\tau}^0_{\;\; j}]_R$ 
(although possibly non--zero) should not make a net contribution to the
correlator. Physically, we may think of this as an expression of the
fact that we do not expect there to be any net bulk momentum transfers
across a radial surface on
the very largest scales, since $\tau^j_{\;\; 0}$ is proportional to a
 well-defined momentum \cite{Amery3}.  This is a much less stringent requirement than
that previously imposed ({\it i.e.} that the functions themselves go to zero beyond
the horizon \cite{pst,tras1}) which is not
expected to be true in most physical situations of interest. 
Our physically motivated and much weaker 
assumption thus allows for a more complete analysis of the
superhorizon behaviour. 

Repeating the analysis, we find that 
\begin{eqnarray}
\int {\rm d} V_{\bf x} \;\; \Phi^0_\beta (\chi_{\bf x}) \tau^{i \;\;\;\; |j}_{\;\;
j|i} = + \int {\rm d} V_{\bf x} \;\; 
\left\{ \left[ \Phi^0_\beta (\chi_{\bf x}) \right]^{,j} \right\}_{|i}
\tau^i_{\;\; j} \; , 
 \label{int-2}
\end{eqnarray}
where we have once again ignored the surface terms. Explicitly, the
only non-trivial surface term is again the radial surface at infinity:
\begin{eqnarray}
\lim_{R \longrightarrow \infty} \frac{\partial \Phi^0_\beta (R)}{\partial R}
\int {\rm d} \Omega_{\bf x} \left[ \tau^i_{\;\; 1} \right]_R \nonumber
\end{eqnarray}
and, this time, we expect there to be no net anisotropic shears
across the radial surface on the largest scales, so that this surface
term averages to zero and may be ignored. Hence, using 
the connection coefficients, as well as the
derivative (\ref{PHI-der}) and recursion (\ref{PHI-recursive})
relations for the $\Phi^0_l$, we see that 
\begin{eqnarray}
\int {\rm d} V_{\bf x} \;\; \Phi^0_{\beta} (\chi_{\bf x}) \tau^{i \;\;\;\;
|j}_{\;\; j|i} = - \int {\rm d} V_{\bf x} \;\; \Phi^0_{\beta} f_K
\left( \tau^2_{\;\; 2} + \tau^3_{\;\; 3} \right)  + \int {\rm d}
V_{\bf x} \;\; \Phi^0_\beta (\chi) \left[ - \tilde{k}^2 + 2 f_K 
\right] \tau^1_{\;\; 1}
\; , 
\label{int-3}
\end{eqnarray}
where
\begin{eqnarray}
f_{K} (\beta, \chi) \equiv \cot^2_K \chi - \beta \cot_K \cot (\beta
\chi) \; , 
\end{eqnarray}
and $\cot_K \chi = \cos_K \chi/\sin_K \chi$ where $\cos_K\chi =  \left\{ \cosh \chi, \chi, \cos \chi \right\}$ for
$\left\{ K < 0,\right.$ $K = 0,$ $\left. K > 0 \right\}$. 

Proceeding in a similar fashion for the ${\bf x} + {\bf y}$ integral, 
multiplying to obtain four double integrals, and then  
reversing the steps that led to (\ref{00-3-correl}) --- {\it i.e.} transforming back
to ${\bf y}$ as the integration variable in the inner integral, and
appealing to  (\ref{phi-split}) --- we may write the $00$--mode as 
\begin{eqnarray}
& & \langle \tau^{i \;\;\;\; |j}_{\;\; j|i} \;\; \tau^{p \;\;\;\;
|q}_{\;\; q|p} \rangle_{00} (\beta) = \nonumber \\
& & \;\;\;\;\;   \langle 
\left[ - \tilde{k}^2 + 2 f_{K} (\beta,\chi_{\bf x}) \right] \tau^1_{\;\; 1} ({\bf
 x}) \;\;  
\left[ - \tilde{k}^2 + 2 f_{K} (\beta,\chi_{\bf x + y}) \right] \tau^1_{\;\; 1}
({\bf x} + {\bf y}) \rangle_{00} (\beta) \nonumber \\
& & \;\;\;\;\;  + \langle \left[ - \tilde{k}^2 + 2 f_{K} (\beta,\chi_{\bf x}) \right] \tau^1_{\;\; 1} ({\bf x}) 
 \;\;  \left[- f_{K} (\beta,\chi_{\bf x + y})\right]  \left( \tau^2_{\;\;
2} + \tau^3_{\;\; 3} \right) ({\bf x} + {\bf y}) \rangle_{00} (\beta) \nonumber \\
& &  \;\;\;\;\;  + \langle - f_{K} (\beta,\chi_{\bf x}) \left( \tau^2_{\;\; 2} + \tau^3_{\;\; 3} \right)
({\bf x}) \;\; \left[ - \tilde{k}^2 + 2 f_{K} (\beta,\chi_{\bf x + y})  \right]
\tau^1_{\;\; 1} ({\bf x} + {\bf y}) \rangle_{00} (\beta) \nonumber \\
& & \;\;\;\;\; + \langle - f_{K} (\beta, \chi_{\bf x}) \left( \tau^2_{\;\;
2} + \tau^3_{\;\; 3} \right) ({\bf x}) \;\;  \left[- f_{K} 
(\beta,\chi_{\bf x + y})\right] \left( \tau^2_{\;\;
2} \;\;\; + \tau^3_{\;\; 3} \right) ({\bf x} + {\bf y}) \rangle_{00} (\beta) \;
. \nonumber \\
& & \label{int-5}
\end{eqnarray}

Each of the four correlators on the right hand side of
(\ref{int-5}) is equal to a double integral similar in form (though,
clearly not in integrand) to (\ref{00-3-correl}).  As remarked earlier, for ${\bf y}$
superhorizon, at least one of ${\bf x}$, ${\bf x + y}$ is superhorizon
also so that we may take the integral over that variable to be
finite.  Effectively, then, we have obtained the transfromed 
two-point correlators 
of the contracted derivatives of the pseudo-stresses---and hence the 
pseudo-energy---in terms of functions of $k$
times the transformed pseudo-stresses.  

\subsection{Flat space}

For the flat space ($K = 0$) case, 
on sufficiently superhorizon scales ($k r \ll 1$) we
may take the limit 
\begin{eqnarray}
f_{K = 0} = 
\frac{1}{r^2} - \frac{k {\rm cot} (kr)}{r} \approx 
\frac{ k^2}{3} \hspace{5mm} {\rm for} \hspace{3mm} kr \ll 1  \; ,
\nonumber 
\end{eqnarray}
to show that the pre-factors $(-k^2 +
2 f_{K=0} )^2$, $- f_{K=0} (-k^2 + 2 f_{K=0})$, $f_{K=0}^2$ are functions of
$k$ only, in the limit $k r \longrightarrow 0$.  Hence, we may pull
these factors out of the (implicit) integrals in (\ref{int-5}) and,
taking the limit to infinity, we obtain
\begin{eqnarray}
\langle \tau^{i \;\;\;\; |j}_{\;\; j|i}  \;\; \tau^{p \;\;\;\;
|q}_{\;\; q|p} \rangle_{00} (k) &=& \frac{k^4}{9}
\left[ \langle \tau^1_{\;\; 1} \;\; \tau^1_{\;\; 1} 
\rangle_{00} (k)  + \langle \tau^1_{\;\; 1} \; \left( \tau^2_{\;\; 2} + \tau^3_{\;\;
3} \right) \rangle_{00} (k)  \right. \nonumber \\
& &  \left. + \langle \left( \tau^2_{\;\; 2} + \tau^3_{\;\; 3} \right) \; 
 \tau^1_{\;\; 1} \rangle_{00} (k)  + \langle  \left( \tau^2_{\;\; 2} + \tau^3_{\;\; 3}
\right)  \; \left(
 \tau^2_{\;\; 2} + \tau^3_{\;\; 3} \right) \rangle_{00} (k)   \right]
\; , \nonumber \\
& & \label{int-6}
\end{eqnarray}
where the correlators on the right hand side of (\ref{int-6}) are
known (see \S \ref{wn-k4sec}) to be proportional to $k^0$ in this
limit.   Hence, appealing to the conservation equations
(\ref{tau-cons}) and noting that time differentiation does not affect
 $\beta$-dependence in our eigenfunction expansions,  
we regain the familiar $k^4$ power spectrum, as the spectrum for the
$00$-mode
\begin{eqnarray}
\langle \tau^{0}_{\;\; 0} ({\bf x}) \;\; \tau^{0}_{\;\;
0} ({\bf x} + {\bf y}) \rangle_{00} (k) \propto k^4 \; . 
\label{K0-k4}
\end{eqnarray} 

\subsection{Open universe} \label{iso-open}

 In this case,  there are three regimes of
interest: supercurvature, 
sub--curvature, and the case in which $\chi$ is near the curvature
scale. On superhorizon, supercurvature scales, $\chi \gg 1$, $\beta
\chi \ll 1$ which implies that $\beta \ll 1/\chi \ll 1$. Hence, we have  that ${\rm \coth}
\chi \approx 1$ and  $\tilde{k} \approx 1$,  where these approximations
are exact in this limit, and about which we have the expansion
\begin{eqnarray}
 f_{K<0} (\beta, \chi) = 1 - \beta
\cot (\beta \chi) = 1 - \frac{1}{\chi} + \frac{\beta^2 \chi}{3} \cdot
\cdot \cdot = O (1) \; , \hspace{10mm} \beta \chi \ll 1 \; , \;\; \chi
\gg 1 \; . \nonumber \\
\label{fkm-bigchi} 
\end{eqnarray}
Substituting (\ref{fkm-bigchi}) into
(\ref{int-5}),  we see that the $00$-mode tends to a
constant on supercurvature, and superhorizon scales.  Explicitly,
making the $\tilde{k} \approx 1$ approximation valid on these superhorizon and
 supercurvature scales, we obtain
\begin{eqnarray}
\langle \tau^{i \;\;\;\; |j}_{\;\; j|i} \;\; \tau^{i \;\;\;\;
|j}_{\;\; j|i}  \rangle_{00} (\beta) &=&
 a_1 \langle \tau^1_{\;\; 1}  \;\;
\tau^1_{\;\; 1}   \rangle_{00} (\beta)  + a_2 \langle \tau^1_{\;\; 1} \;
 \left( \tau^2_{\;\; 2} + \tau^3_{\;\; 3}
 \right)  \rangle_{00} (\beta)  \nonumber \\
& & + a_3 \langle \left( \tau^2_{\;\; 2} +
 \tau^3_{\;\; 3} \right) \; \tau^1_{\;\; 1}
 \rangle_{00} (\beta)  + a_4\langle \left( \tau^2_{\;\; 2} +
 \tau^3_{\;\; 3} \right)   \;\left( \tau^2_{\;\; 2} + \tau^3_{\;\; 3}
 \right)  \rangle_{00} (\beta) \; , 
\nonumber \\
  \hspace{30mm} {\rm for } & & \hspace{10mm}  
\chi_{{\bf x},{\bf y}} \gg 1,  \hspace{5mm}
\beta \chi_{{\bf x},{\bf y}} \ll 1 \; , \nonumber 
\end{eqnarray}  
where the $a_i$ are constant factors. 
Recalling the equations of motion (\ref{tau-cons}) for the energy-momentum
pseudo-tensor, and that the correlator
between two arbitrary functions goes as $\beta^0$, we see that on very
supercurvature and very superhorizon scales, the correlators behave as:  
\begin{eqnarray}
\langle \tau^0_{\;\; 0}
\;\; \tau^0_{\;\; 0} \rangle_{00} \; \; \propto \; \; 
\beta^0 \; . \label{Kn0-r1} 
\end{eqnarray}

Turning now to the other asymptote, we consider the case in which the
horizon is very much less than the curvature scale. Then, on
 superhorizon, and very much subcurvature scales, we regain the
familiar $k^4$ power spectrum, up to a constant factor.  Using the
series expansions appropriate
for $\chi \ll 1$ and $\beta \chi \ll 1$, one easily finds that 
\begin{eqnarray}
f_{K<0} = \frac{\beta^2 + 1}{3} = \frac{\tilde{k}^2}{3} \; , \hspace{10mm}
\beta \chi \ll 1 \; , \chi \ll 1 \; . 
\label{fkm-litchi}
\end{eqnarray}
This has the same form as the superhorizon limit of $f_{K=0}$.
However, one should recall that, while $\beta$ (for the $K < 0$ case) and
$k$ (for $K = 0$) have similar spectra, their meanings are different
in the two spacetimes --- see equation (\ref{betak-eqn}). In particular, we
have, for the open universe with $\chi \ll 1$, two cases: $\beta < 1$ and $\beta > 
1$. These correspond to mode wavelengths approximately equal to  and smaller than
the curvature horizon, respectively. In the former case, $f_{K<0}
\approx 1/3$ and $\tilde{k} \approx 1$  so  that we have the 
result (\ref{Kn0-r1}) once more.  In the latter, $f_{K<0} \approx \beta^2/3$, for
large $\beta$, in which case we also have that $\tilde{k}^2 \approx
\beta^2$. Then, recalling the normalisation of  $k$, we have the local
(subcurvature) superhorizon limit: 
\begin{eqnarray}
 f_{K<0} (\chi, \beta) \approx
\frac{1}{|K|} f_{K=0} (\tilde{k}, r) \; . \nonumber
\end{eqnarray}
Recalling the conservation equations (\ref{tau-cons}), this means that 
\begin{eqnarray}
\langle \tau^0_{\;\; 0} ({\bf x})  \;\; \tau^0_{\;\; 0} ({\bf x} + {\bf
y}) \rangle_{00} (\beta) ~\propto~ \tilde{k}^4 = \frac{k^4}{|K|^2}\approx \beta^4, \hspace{8mm} \chi_{{\bf x},{\bf y}} \ll 1, \hspace{5mm}
\beta \chi_{{\bf x},{\bf y}} \ll 1 \; , \label{Kn0-r2}
\end{eqnarray}  
where $k$ is the flat space Fourier wavenumber. 

Having dealt with both the extreme cases of local flatness and
 curvature domination, there remains the interesting
question of the behaviour near the curvature scale. We therefore would
like to understand how the $k^4$ superhorizon behaviour on
sub--curvature scales becomes $\beta^0 = k^0$ on very super--curvature
scales, and the ranges within which these approximations are valid.   

Away from the asymptotes dealt with
above, $f_{K<0}$ is a function of the radial coordinate, and therefore
is integrated over (twice) in equation (\ref{int-5}).  This means that
we may only truly understand the effect of the factors in this
equation by explicitly calculating the integral, which requires
that we specify the pseudo--stresses.   As a first
approximation, however, we may proceed to obtain some understanding of
the superhorizon behaviour of generic causal perturbations by
assuming that the corrections are small enough for us to make the
approximation
\begin{eqnarray}
< \tau^0_{\;\; 0}({\bf x}) \;\; \tau^0_{\;\; 0}({\bf x + y}) >_{00}~
\approx ~(P1 + P2 + P3 + P4) A_{00} \; , \label{ch4-app1}
\end{eqnarray}
where 
\begin{eqnarray} 
\begin{array}{ll}
P1 = ( -\tilde{k}^2 + 2 f_{K<0} )^2  \; , \hspace{15mm} &  P2 = -
f_{K<0} ( -\tilde{k}^2 + 2
f_{K<0}) \; , \\
P3 = - (-\tilde{k}^2 + 2 f_{K<0}) f_{K<0} \; , \hspace{15mm} & P4 = f_{K<0}^2
\; , 
\end{array}
\nonumber
\end{eqnarray}
for some given $\chi$, and where we have assumed that the correlators
of the $\tau^i_{\;\; j}$ are well modelled by window functions of the
form used in \S \ref{HS-k4sec}.  This assumption amounts to ignoring
long range correlations, since we assume that $\chi_{\bf x} \approx
\chi_{\bf x + y}$.  Hence this approximation is valid for $\chi_{\bf y}$
small, that is, in the very early universe.  

\begin{figure}
\leavevmode\epsfxsize=7.5cm \epsfbox{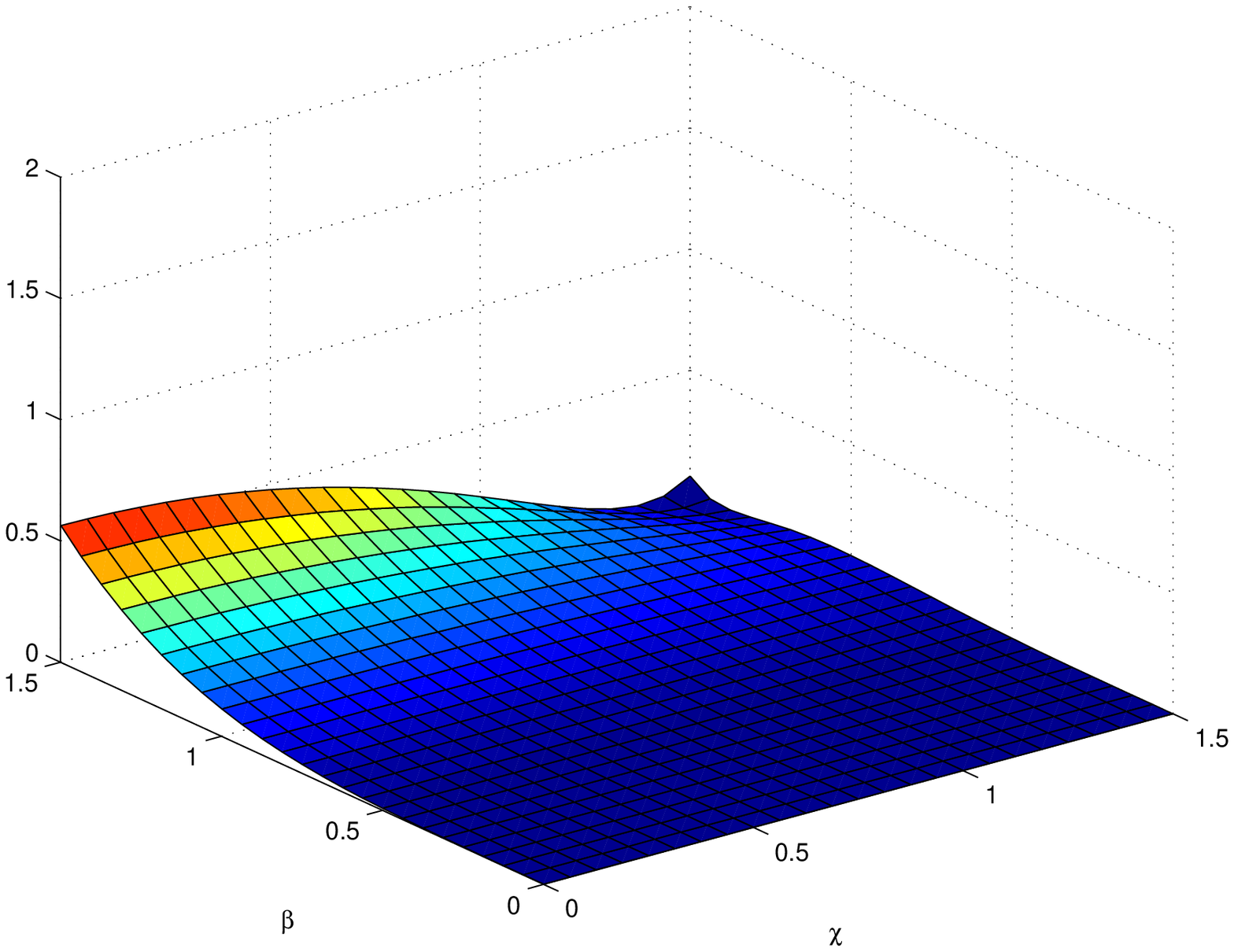}
\leavevmode\epsfxsize=7.5cm \epsfbox{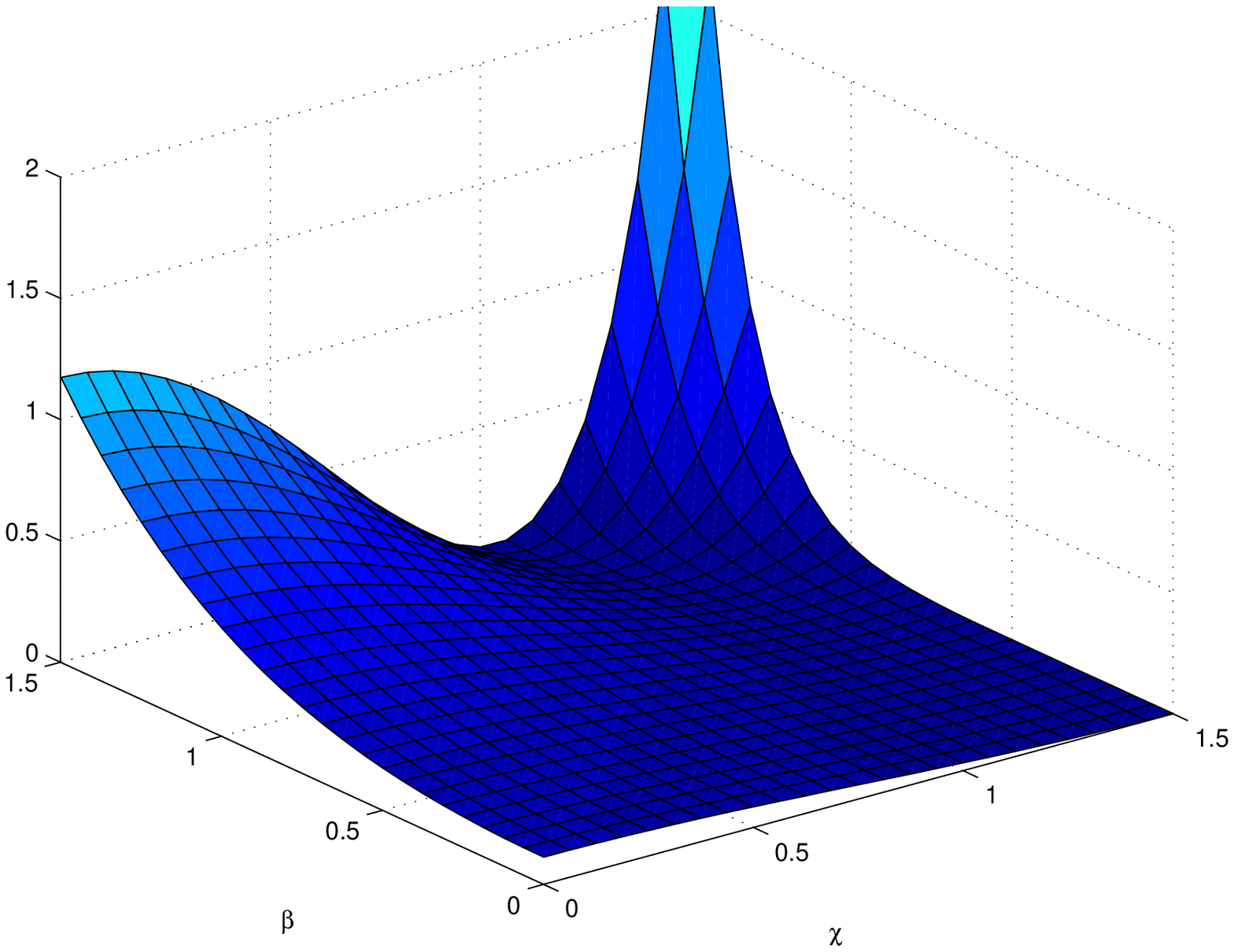}\\ 
  \caption[]
  {P1: K = 0 (LHS), and K = -1 (RHS)}
  \label{fig-s1}
\end{figure}

\begin{figure}
\leavevmode\epsfxsize=7.5cm \epsfbox{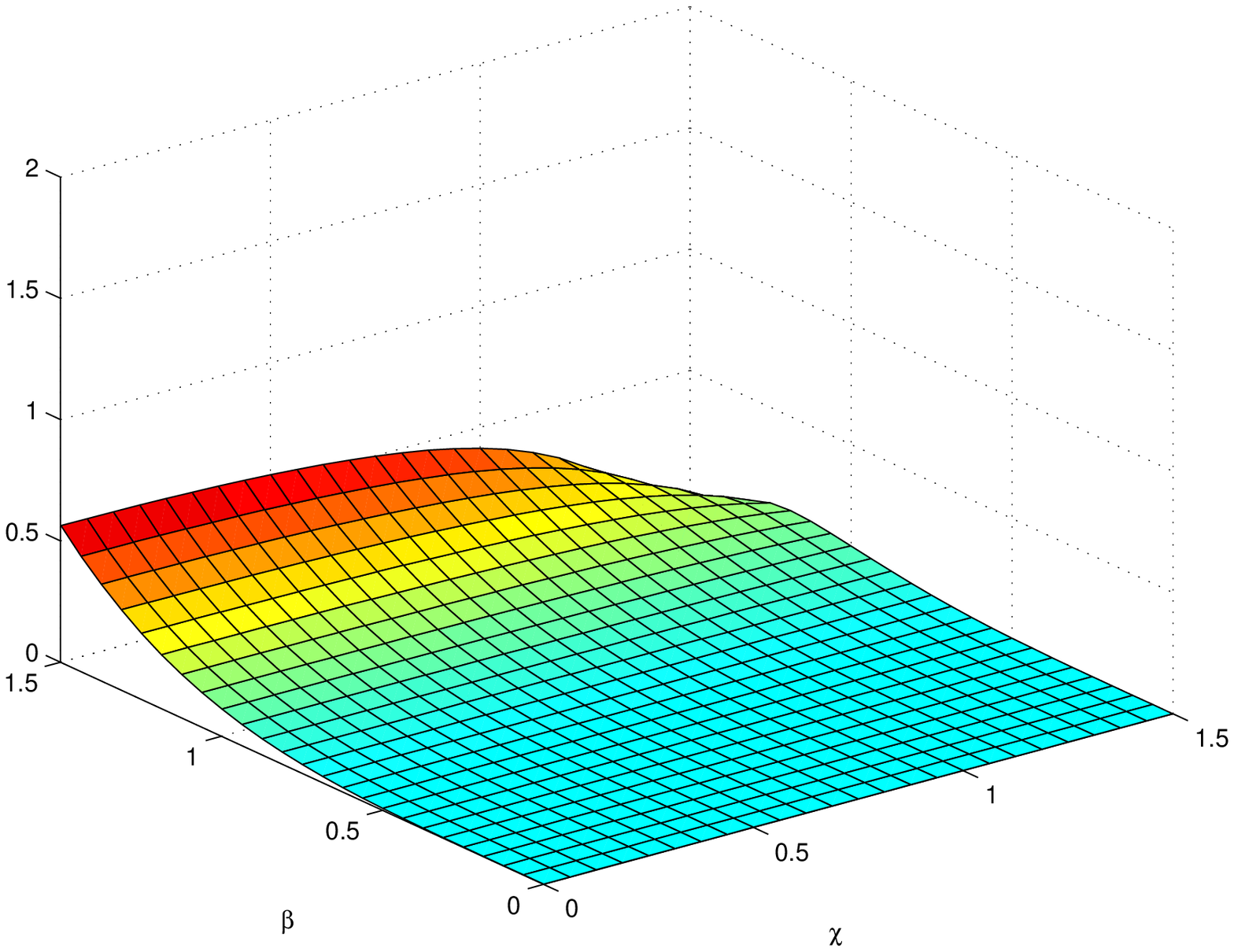}
\leavevmode\epsfxsize=7.5cm \epsfbox{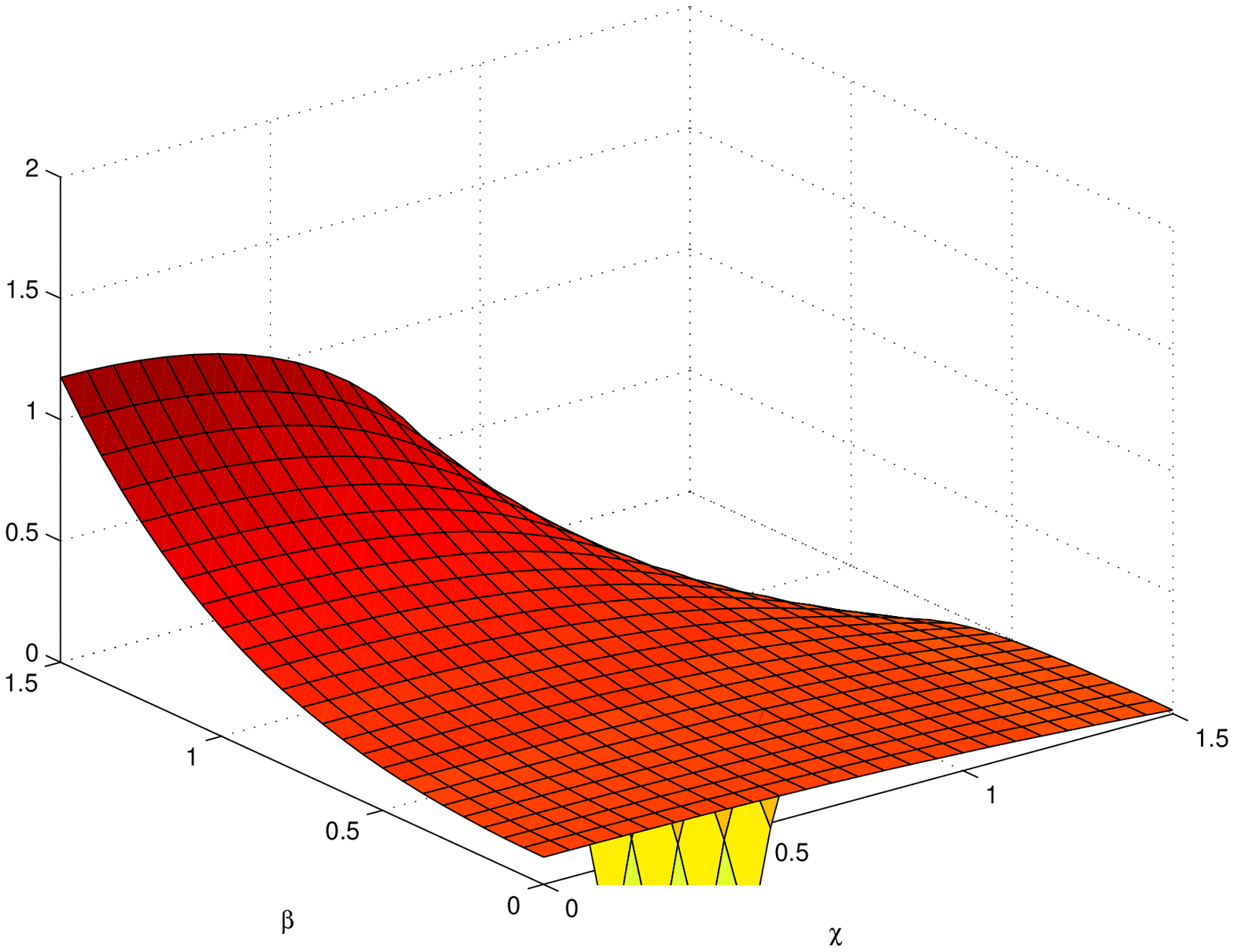}\\ 
  \caption[]
  {P2 = P3 (in this limit): K = 0 (LHS), K = -1 (RHS)}
  \label{fig-s2}
\end{figure}

\begin{figure}
\leavevmode\epsfxsize=7.5cm \epsfbox{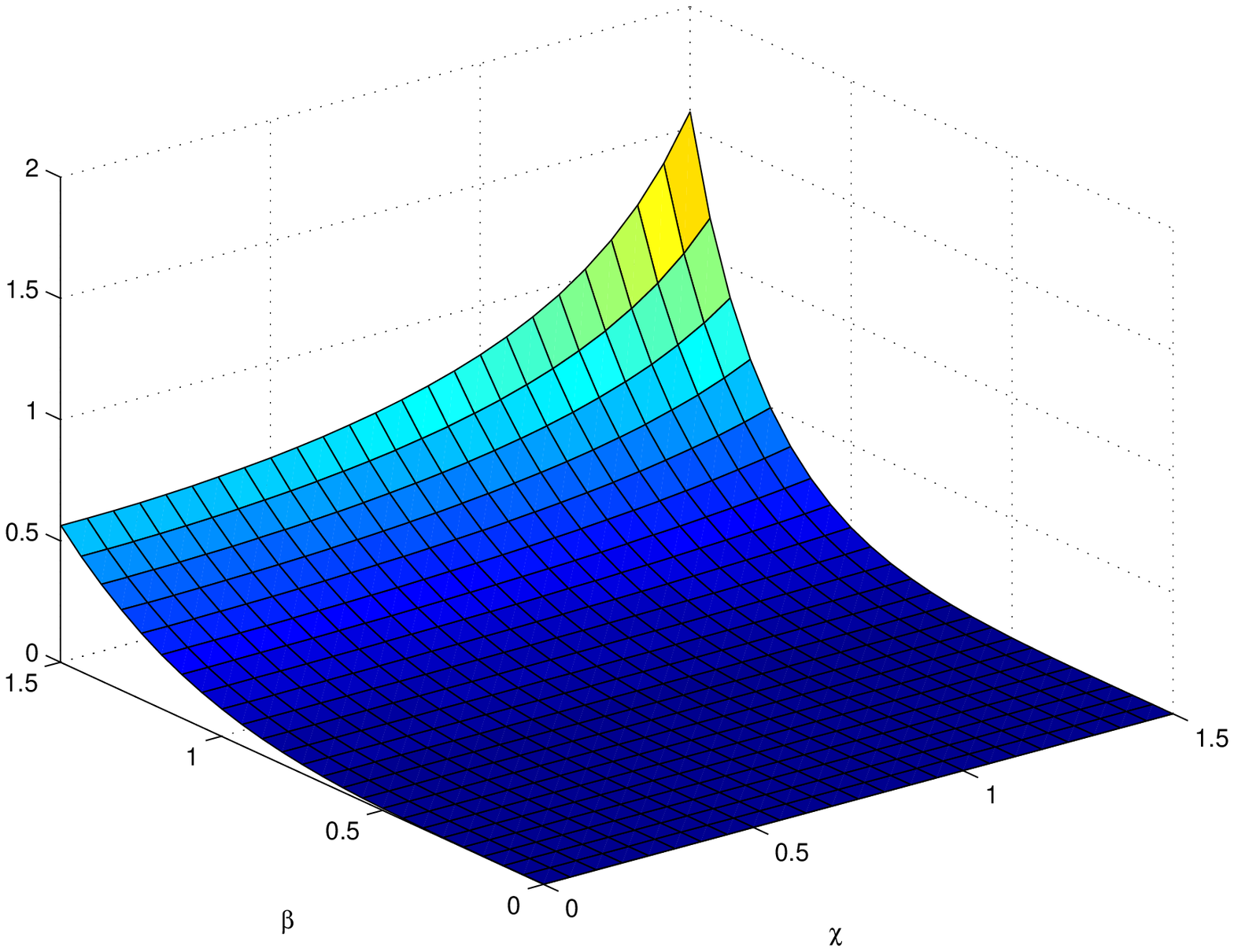}
\leavevmode\epsfxsize=7.5cm \epsfbox{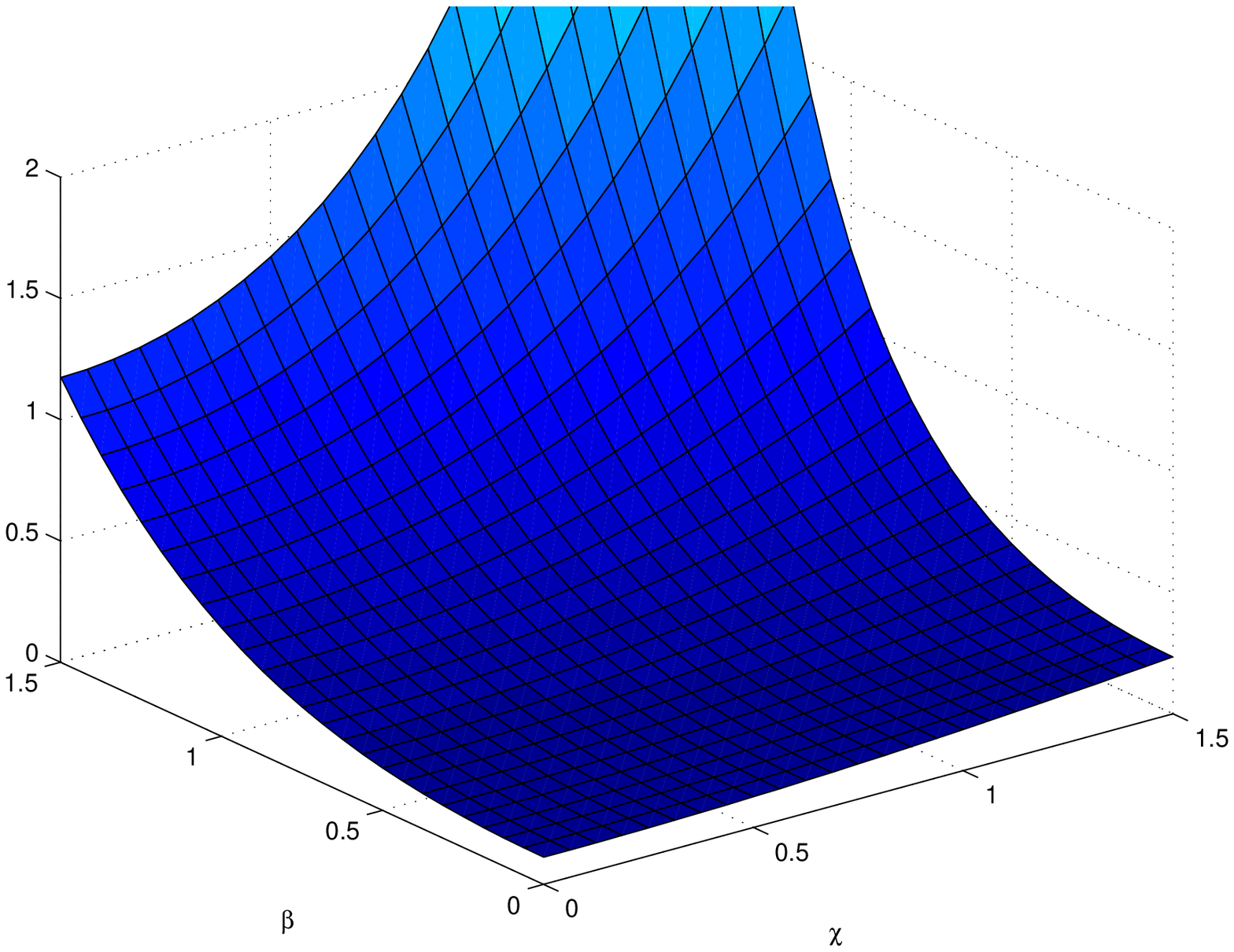}\\  
  \caption[]
  {P4: K = 0 (LHS), K = -1 (RHS)}
  \label{fig-s3}
\end{figure}

We provide surface
plots for these pre-factors in Figures
\ref{fig-s1}--\ref{fig-s3}, from which we can clearly see how our
approximations are valid only for $\chi < 0.5$, and break down as we
move into the large $\chi$ and large $\beta$ regime, doing so much
earlier than in the $K = 0$ scenario.  This is a geometric effect due
to the fact that the majority of the volume in a hyperbolic space
is located near the radial surface, so that a large `hyper-sphere' approximates
a `hyper-shell'.  This effect is most noticible on supercurvature
scales. However, one can calculate that the ratio of hyperbolic volume to
surface area rapidly diverges from that for a spherical geometry,
beginning at around $\chi \sim 0.5$. On
the other hand, the steep slope in the prefactor $P_1$ for large
$\beta$ and $\chi < 0.4$ is due to the appearance of a constant
(curvature) term 
in the (repeated) $\tilde{k}$ factors in this prefactor. However, as 
$\chi$ increases, so does
$f_{K<0}$, first cancelling, and then overriding this effect.
Along the
$\beta$ axis in each plot ({\it i.e.} for $\chi \sim 0$), we see that all
the pre-factors go as $k^4 / 9$ for $K = 0$ and as $(\beta^2 + 1)^2
/9$ for $K = -1$. Along the $\chi$ axis ($\beta \sim 0$), all the
pre-factors vanish for
$K = 0$.  However, for the $K = -1$ case, this is not the
case, reflecting the small constant (curvature) term.  

\begin{figure}
\leavevmode\epsfxsize=7.5cm \epsfbox{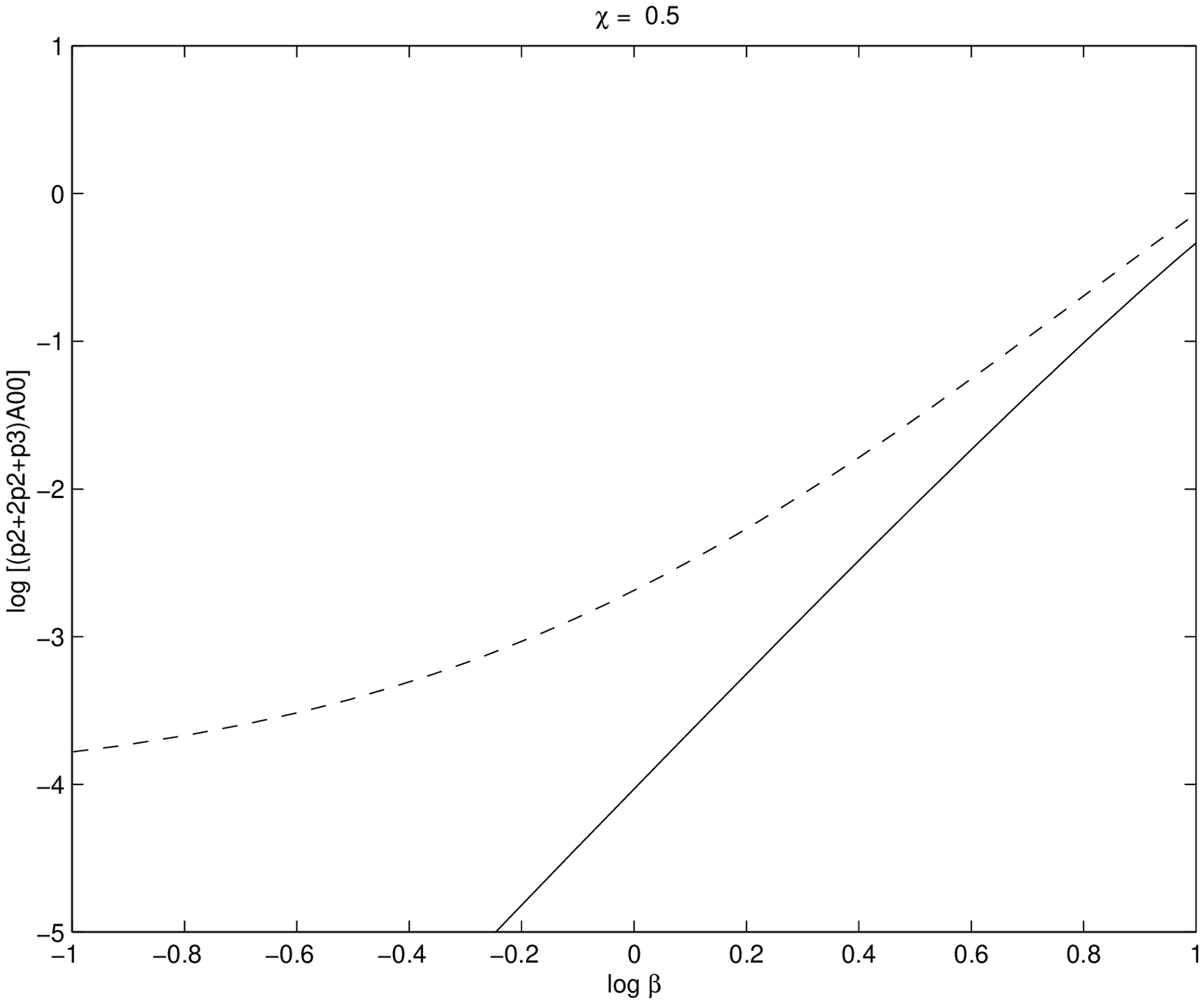}
\leavevmode\epsfxsize=7.5cm \epsfbox{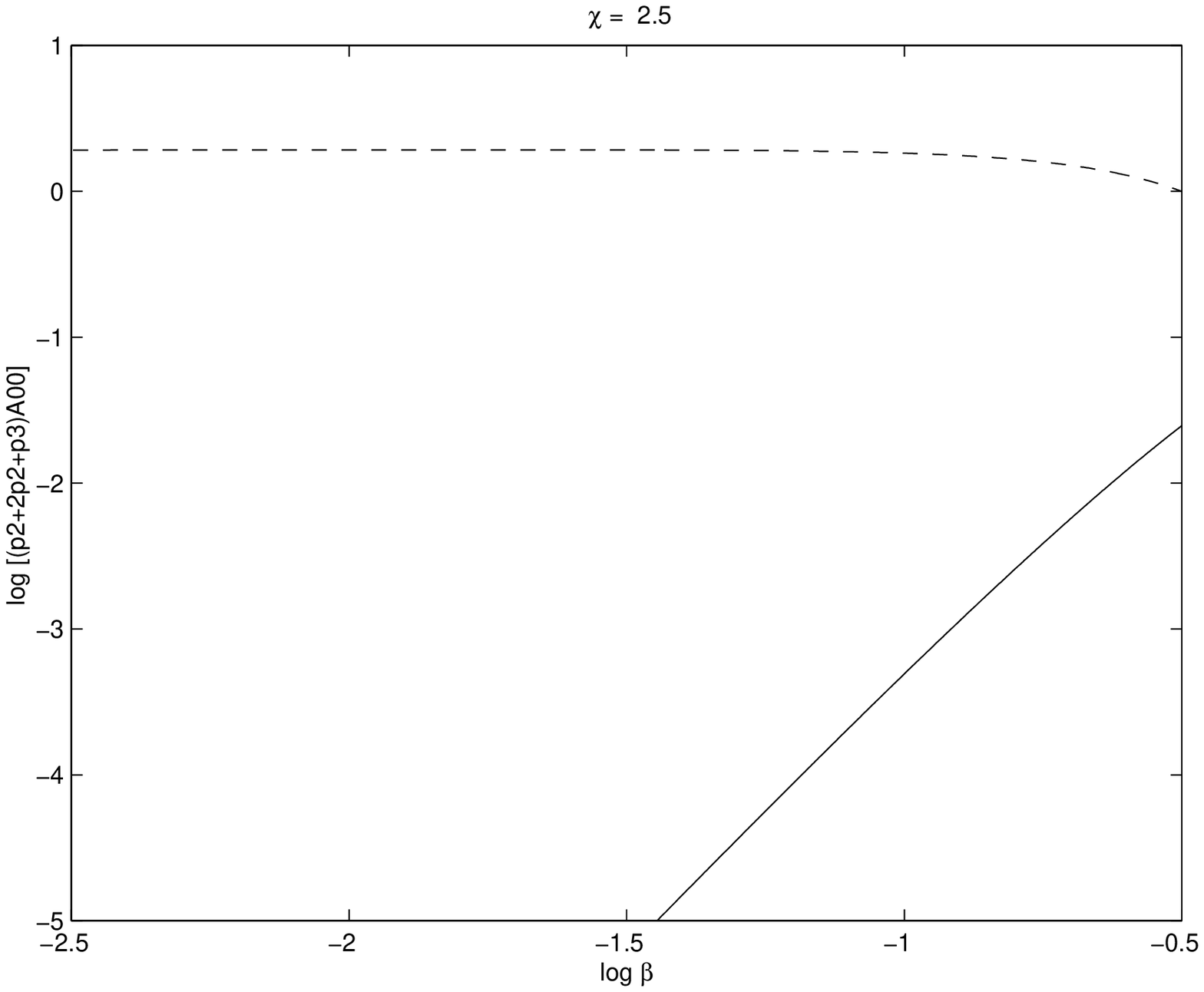}\\ 
  \caption[]
  {Log (P1 + 2P2 + P3)A00 for K = 0 (-) and K = -1 (- - -)}
  \label{fig-z1}
\end{figure}

Computing
 $\langle \tau^0_{\;\; 0}  \;\; \tau^0_{\;\; 0} \rangle$ using the
approximation (\ref{ch4-app1}) we find the curves plotted in Figure
\ref{fig-z1}.  These show clearly how the $k^4$ behaviour on
superhorizon scales is flattened out, for small $\beta$, by the intrinsic curvature. We
 note that this does not mean that we expect the real space
 correlators to similarly go to a constant on these scales because, of course,
 they  vanish by causality. Integrating (\ref{ch4-app1}) over all $\beta$, with the
 kernel $\Phi^0_\beta$,  the constant term is
 suppressed by the measure ${\rm d} \mu (\beta) = \beta^2 {\rm d}
 \beta$. Similarly, calculating the variance 
\begin{eqnarray}
\sigma^2 = \int_0^{\infty} {\rm d} \mu (\beta) \langle \tau^0_{\;\; 0} \;\;
\tau^0_{\;\; 0} \rangle_{00} (\beta)  =  \int_0^{1} {\rm d} \mu (\beta) \langle \tau^0_{\;\; 0} \;\;
\tau^0_{\;\; 0} \rangle_{00} (\beta) + \int_1^{\infty} {\rm d} \mu (\beta) \langle \tau^0_{\;\; 0} \;\;
\tau^0_{\;\; 0} \rangle_{00} (\beta) \; , \nonumber 
\end{eqnarray}
and breaking the integral
 over all $\beta$ into supercurvature and subcurvature
  parts, we see that the second integral is procedurally 
 the same as for the flat space case, for both sub- and
 superhorizon scales, while the first
 (containing the constant power spectrum constribution) is
 suppressed by the $\beta^2$ factor in the measure.

\subsection{Closed universe} \label{isop-k4sec}

For the closed universe we have the prefactors determined by 
\begin{eqnarray}
f_{K>0} (\beta, \chi) = {\rm cot}^2 \chi - \beta \cot \chi \cot (\beta
\chi) \; . \label{fK>0-1}
\end{eqnarray}
The difference between the open and closed universe cases is
 quantified by the differences between the behaviour of the function
 (\ref{fK>0-1}) 
 and that of $f_{K<0}$. To begin with, the eigenspectrum for 
 $\beta$ is discrete, and the lowest non-gauge mode is $\beta =
3$ so that the superhorizon limit $\beta \chi_H \ll 1$ can only be attained
 for $\chi_H \ll 1/3$.  Of course, for $\chi_H \gg 1$ we encounter the 
periodic properties the closed universe: 
one cannot have modes that are both superhorizon and
 supercurvature.   

If we take $\chi \ll 1$ and $\beta \chi \ll 1$, then we consider the 
subcurvature yet superhorizon case.  Here, we note
that $\cot \chi \approx 1/ \chi - \chi/3$, wo we have 
\begin{eqnarray}
f_{K>0} = \frac{\beta^2 - 1}{3} = \frac{\tilde{k}^2}{3} \; , \hspace{10mm}
\beta \chi \ll 1, \;\;\; \chi \ll 1 \; , \label{fK>0-2}
\end{eqnarray}
and we regain the flat space results in a similar fashion to the open universe 
case. Hence, an analysis similar to that for the open
 universe shows that we once again find the local limit
\begin{eqnarray}
\langle \tau^0_{\;\; 0} ({\bf x}) \tau^0_{\;\; 0} ({\bf x + y})
 \rangle \propto \frac{\tilde{k}^4}{9} \label{closed-2}
\end{eqnarray}
where $\tilde{k}^4 = (\beta^2 - 1)^2 \approx k^4 / |K|^2$ for large
$\beta$, with $k$ being the flat space Fourier wavenumber. 

\section{Discussion} \label{INITCON} 

It would be of great physical interest if the consequences of this 
superhorizon behaviour were within the range of observational searches
because it would test  `active' theories for structure formation, and 
possibly probe the curvature of the universe.  The cosmic microwave sky
may provide the best prospect, however, for the CMB temperature spectrum 
the superhorizon `fall-off' expected on the very largest scales will be
confused by cosmic variance.  The most likely signature would be through 
the CMB polarization spectrum because it represents a primordial snapshot
of the universe at recombination which includes about $10^5$ different
horizon volumes.  The lack of spatial correlations on scales corresponding to 
the horizon size at that time (about $2^\circ$), and the fall-off in 
their transformed counterparts, should be observationally testable. 
Previous analysis of this possibility \cite{SZ}, however, has only 
considered scalar modes, so we comment on the possible impact of vector
and tensor modes here.  We leave discussion of the observational imprint
of this superhorizon behaviour for a future publication \cite{Amery5}, 
concentrating instead on the consistent implementation of initial 
conditions for numerical simulations.   

\subsection{SVT decomposition and superhorizon coherence} \label{SHCOH}

The decomposition of a tensor perturbation into its scalar, vector and tensor parts is not 
 without its subtleties \cite{vs}, since the projection operators yielding the scalar, vector and 
 tensor parts are not, in general, analytic. 
The spatial distribution of the separate geometric parts may therefore be 
 significantly different from the physical distribution of the entire object, so that the 
 decomposed parts of a causal tensor object can manifest acausal features. The transformed 
 variables will then in turn exhibit non-analytic behaviour \cite{durrer}, complicating the 
 analysis of their superhorizon behaviour.  However, causality is merely hidden, and, since the 
 arguments presented in this paper for the (scalar) energy density perturbation are (trivially) 
 phrased in terms of full tensors, they  escape this difficulty.  

On the other hand, it is common to employ the assumption of coherence, in which case  one regains 
 analyticity \cite{durrer}, all the decomposed parts of the 
 perturbation are causal, and one may obtain the power spectra by  
 simple power counting.  This allows us to write down the auto-correlators for both the   
 scalar and the vector parts of  the pseudo-tensor $\tau^\mu_{\;\; \nu}$, under the influence of 
 energy-momentum conservation.   

If we decompose the pseudo-tensor $\tau^\mu_{\;\; \nu}$ as in
(\ref{met-decomp}), we may write the energy-momentum equations
(\ref{tau-cons}) as 
\begin{eqnarray}
\tau_{IV} &=& \frac{\dot{\tau}_S}{{k}} \; , \hspace{20mm} 
k \dot{\tau}_{IV} = 2 {k}^2 \left[ \tau_L + 2 \left( \frac{K}{{k}^2} - \frac{1}{3} \right)
\tau_T \right]  = \ddot{\tau}_S \; , \nonumber \\
\tau^{(\pm 1)}_{IT} &=& - 2 {k} [ {k}^2 - 2K ]^{-1} \dot{\tau}^{(\pm 1)}_V =
\frac{\ddot{h}^{(\pm 1)}_V}{\kappa} \; , \label{T-DEF2-3}
\end{eqnarray}
where we have used the first equation to obtain the final equality in
the second.  Assuming coherence, we find  that the 
 scalar ($\tau_S$) and vector ($\tau_{IV}$, $\tau_V$) degrees of
freedom have auto-correlators that go as 
\begin{eqnarray}
\langle |\tau_S|^2 \rangle (\beta) &~~\propto~~& \tilde{k}^2 \langle |\tau_{IV}|^2  \rangle (\beta) \; , \nonumber \\
\langle |\tau_{IV}|^2 \rangle (\beta) &~~\propto~~& \tilde{k}^2 \left( 4 \langle |\tau_{L}|^2 \rangle (\beta) - \frac{16}{3} \langle \tau_L \tau_T^* \rangle (\beta) + \frac{16}{9} \langle |\tau_T|^2 \rangle (\beta) \right) \nonumber \\
 &~~~~~~+& 16 K \left( \langle | \tau_L|^2 \rangle - \frac{2}{3} \langle | \tau_T|^2\rangle (\beta) \right) 
+ 16 \frac{K^2}{\tilde{k}^2} \langle | \tau_T |^2\rangle (\beta) \; , \nonumber \\
\langle |\tau^{(\pm 1)}_V |^2 \rangle (\beta) &~~\propto~~& \left( \frac{1}{4} \tilde{k}^2 - K + \frac{K^2}{\tilde{k}^2} \right) \langle |\tau^{(\pm 1)}_{IT} |^2 \rangle (\beta) \; , \label{ApCoSh}
\end{eqnarray}
on superhorizon scales. Moreover, the correlators for $\tau_L$, $\tau_T$, and $\tau_{IT}$  have white noise
  superhorizon behaviour (refer to \S \ref{wn-k4sec} and \cite{durrer}) so we see that in the local limit 
($\chi << 1$, $\beta >> 1$) we regain the flat space results, with the scalar
power spectrum $\propto \tilde{k}^4$, while those of the vectors
is proportional to $\tilde{k}^2$.  In the
supercurvature limit, they all tend to a constant. Note that the correlators in (\ref{ApCoSh}) 
 do not diverge as $\beta \longrightarrow 0$, since $\tilde{k}$ tends to $1$ in this limit. 

The results (\ref{ApCoSh}) are more general than one might suppose upon first inspection, as  
 an arbitrary decoherent source may be written as a `decoherent sum' of 
 coherent sources, each of which will then be amenable to the above arguments \cite{turok1}.  
 Moreover, 
 we believe that it is plausible to expect approximate coherence on the  
 very largest scales, since the eigenfunctions used to diagonalise the `decoherent sum' are 
 likely to either all tend to white noise on these scales, or to be dominated by just one element 
 in the sum --- see also ref. \cite{Amery5}.  This notion  has been ventured before as an
 hypothesis \cite{DS} for scaling sources, and is supported by both numerical
results \cite{KD}, and by the observation that, for causal sources,
modes with wavelengths much larger than the horizon
cannot be significantly out of phase with one another.  We shall consider this
point in more detail elsewhere \cite{Amery5}.

\subsection{Matching conditions and initial conditions}

Finally, we wish to propose a set of initial conditions which 
would find useful application in numerical simulations of 
causal perturbation models, such as cosmic defects.  We 
shall assume, as is usual, that we are dealing primarily 
 with the evolution of transformed quantities.
By considered an instantaneous phase transition early in the
universe as a first approximation to a model for cosmic defects
``switching on'',  and employing  
 a gauge in which constant energy and constant time surfaces
coincide,  matching conditions can be used to show that 
 there exists an entire class of objects which are  continuous across
the transition  and are related by gauge
transformation to our pseudo-tensor components.  In a previous paper
(ref. \cite{Amery3}), we used the notion of
compensation together with a particular gauge specification to remove this  redundancy  such
that the  $\tau_S$ (pseudo-energy) and $\tau^{(\pm 1)}_V$
(divergenceless vector) components of our
generalized pseudo-tensor have this property:
\begin{eqnarray}
[\tau_S ]_{\pm} = 0 \; , \hspace{15mm} [ \tau_V^{\pm 1}]_{\pm} = 0 \,. \nonumber 
\end{eqnarray}
While we emphasise that this matching depends on a special gauge choice, it is one which 
should be 
widely applicable in many physical situations, for example, a defect-forming transition 
\cite{Amery3}.

For a universe which
was unperturbed (and hence homogeneous and isotropic) prior to
the transition, we may then take  $\tau_S = 0 = \tau^{(\pm 1)}_V$  as natural initial
conditions.  This result is consistent on all
scales and is especially appropriate for suppressed superhorizon modes. 
The results of \S \ref{SEC-SHiso} confirm this conclusion for the
density perturbation tracked by $\tau_S$, and the previous section argues that this
conclusion may be extended also to the `induced vector' mode
$\tau_{IV}$ and the true vector modes $\tau_V^{\pm 1}$.  
Hence, if one wishes to set initial conditions for perturbations initially 
outside the horizon, one may
consistently  set both the pseudo-energy $\tau^0_{\;\;0}$ and the pseudo-momentum
$\tau^0_{\;\;i}$ to zero.  If one wishes to
include long wavelength modes that are affected by curvature, then one can 
proceed in a similar fashion.  Although the transformed correlators become 
constant on these scales, at early times they are sufficiently outside the
horizon for their amplitude to be strongly suppressed.   Therefore, the 
matching conditions we propose as initial conditions will also be 
applicable and self-consistent for a curved universe.

\section*{Acknowledgements}

We are grateful for useful discussions with Rob Crittenden, 
Martin Landriau, Neil Turok,
and Proty Wu.  GA acknowledges the support of the
Cambridge Commonwealth Trust; ORS; Trinity Hall; Cecil Renaud Educational 
and Charitable Trust.  This work was supported by PPARC grant no.  
PPA/G/O/1999/00603.

\appendix
\section{Eigenfunctions of the Laplacian}

The scalar eigenfunctions are given by solutions to (\ref{eqn-Helm})
with $|m| = 0$ ({\it c.f.}, with a slightly different normalisation, ref. \cite{as}): 
\begin{eqnarray}
\zeta_{\beta l m}
({\bf x})
= \sqrt{\frac{2}{\pi}}\Phi^\beta_l (y)  Y^m_l (\theta, \phi) \; . 
\label{sca-emode1}
\end{eqnarray} 
Here the angular dependence is contained in the spherical harmonics
which are given
by 
$Y_l^m (\theta, \phi) = \sqrt{\frac{2l + 1}{4 \pi} \frac{(l -
m)!}{(l + m)!}} P_l^m ({\rm cos} \theta ) {\rm e}^{i m \phi}$  
(where $P_l^m$ is an associated Legendre function), while the radial
function $\Phi^\beta_l$ is 
\begin{eqnarray}
\Phi^l_{\beta} = \left\{ \begin{array}{ll}
\left( \frac{\pi N^l_{\beta}}{2 \beta^2 {\rm sinh} \chi}
\right)^{\frac{1}{2}}  P^{- \frac{1}{2} - l}_{- \frac{1}{2} + \imath
\beta} ( {\rm cosh} \chi ) & \hspace{10mm} K < 0 \\
j_l (\beta \chi ) = j_l (kr) & \hspace{10mm} K = 0 \\
\left( \frac{\pi M^l_{\beta}}{2 \beta^2 {\rm sin} \chi}
\right)^{\frac{1}{2}}  P^{- \frac{1}{2} - l}_{- \frac{1}{2} + \beta} (
{\rm cos} \chi ) & \hspace{10mm} K = > 0 
\end{array}
\right. \; , 
\label{Phi-soln}
\end{eqnarray}
with $N^l_{\beta} = \Pi^l_{n = 0} ( \beta^2 + n^2)$ and  
$M^l_{\beta} = \Pi^l_{n = 0} (\beta^2 - n^2)$.  By Taylor expanding
about $\beta \chi$ small, we may show that $\Phi^l_\beta \sim j_l
(kr)$ in the limit $y \longrightarrow 0$, $\beta \longrightarrow
\infty$. 

The basis $\{ \zeta_{\beta l m} \}$ is orthonormal as well
as complete, and is normalised according to 
\begin{eqnarray}
\int_S {\rm d} V \zeta^*_{\beta^\prime l^\prime m^\prime} \zeta_{\beta
l m} = \delta (\beta^\prime, \beta) \delta_{l^\prime l}
\delta_{m^\prime m} \; , 
\label{normcond}
\end{eqnarray}
where $\delta (\beta^\prime \beta)$ is the delta function with respect
to the measure $\mu (\beta)$: 
\begin{eqnarray}
\begin{array}{lll}
\int {\rm d} \mu (\beta) = \int_0^\infty \beta^2 {\rm d} \beta \; , & 
\hspace{10mm} \delta (\beta, \beta^\prime) = \frac{1}{\beta^2} \delta
(\beta - \beta^\prime) \; , & \hspace{10mm} {\rm for} \;\; K \leq 0 \;
, \\
\int {\rm d} \mu (\beta) = \sum^{\infty}_{\beta = 3} \beta^2 \; , & 
\hspace{10mm} \delta (\beta, \beta^\prime) = \frac{1}{\beta^2}
\delta_{\beta \beta^\prime} \; , & \hspace{10mm} {\rm for} \;\; K > 0 \;
. 
\end{array}
\nonumber 
\end{eqnarray}

The $l = 0$ and $l = 1$ radial modes are given \cite{as} by the exact formulae
 (rewritten for a continuous range of $K$ values): 
\begin{eqnarray}
\Phi^0_{\beta} = \frac{ {\rm sin} (\beta \chi)}{\beta} \times \left\{
\begin{array}{ll}
\frac{1}{{\rm sinh} \chi} & \hspace{10mm} K < 0 \\
\frac{1}{\chi} & \hspace{10mm} K = 0 \\
\frac{1}{{\rm sin} \chi} & \hspace{10mm} K > 0
\end{array} \right. \; , 
\label{Phi-l0} \\
\Phi^1_\beta = \Phi^0_\beta \times \left\{
\begin{array}{ll}
\left( \beta^2 + 1 \right)^{- \frac{1}{2}} \left[
\coth \chi - \beta \cot (\beta \chi ) \right] & K < 0 \\
\frac{1}{\beta} \left[ \frac{1}{\chi} - \beta \cot (\beta \chi) \right] 
& K = 0 \\
\left( \beta^2 - 1 \right)^{- \frac{1}{2}} \left[
\cot \chi - \beta \cot (\beta \chi ) \right] & K > 0 
\end{array} \right. \; . \label{Phi-l1}
\end{eqnarray}

The radial function $\Phi^l_\beta$ may be recursively calculated using
the exact formulae (\ref{Phi-l0}), (\ref{Phi-l1}) and the relations:
\begin{eqnarray}
\Phi^l_\beta = \left\{ 
\begin{array}{ll}
\left( \beta^2 + l^2 \right)^{- \frac{1}{2}} \left[ (2l - 1) (\coth \chi)
\Phi^{l-1}_\beta - \left[ \beta^2 + (l - 1)^2 \right]^{\frac{1}{2}} 
\Phi^{l-2}_\beta \right] & K < 0 \\
\frac{1}{\beta} \left[ (2l - 1) \frac{1}{\chi} \Phi^{l-1}_\beta - 
\beta \Phi^{l-2}_\beta \right] & K = 0 \\
\left( \beta^2 - l^2 \right)^{- \frac{1}{2}} \left[ (2l - 1) (\cot \chi)
\Phi^{l-1}_\beta - \left[ \beta^2 - (l - 1)^2 \right]^{\frac{1}{2}} 
\Phi^{l-2}_\beta \right] & K > 0 
\end{array} \right. \; , \label{PHI-recursive}
\end{eqnarray}
while the derivative with respect to the radial coordinate may be found
using
\begin{eqnarray}
\frac{{\rm d}}{{\rm d} \chi} \Phi^l_\beta = \left\{
\begin{array}{ll} 
l ( \coth \chi ) \Phi^l_\beta - \left[ \beta^2 + ( l + 1 )^2
\right]^{\frac{1}{2}} \Phi^{l+1}_\beta & K < 0 \\
\frac{l}{\chi} \Phi^l_\beta - \beta \Phi^{l+1}_\beta & K = 0 \\
l ( \cot \chi ) \Phi^l_\beta - \left[ \beta^2 - ( l + 1 )^2
\right]^{\frac{1}{2}} \Phi^{l+1}_\beta & K > 0 
\end{array} \right. \; . 
\label{PHI-der}
\end{eqnarray}
Using (\ref{PHI-der}) and 
(\ref{PHI-recursive}) for $\Phi^1_\beta$ and $\Phi^2_\beta$
respectively, one may obtain the relations:
\begin{eqnarray}
\begin{array}{ll}
\frac{{\rm d}}{{\rm d} \chi} \left[ \frac{{\rm sinh}^2 \chi}{\sqrt{\beta^2 +
1}} \Phi^1_\beta (\chi) \right] = {\rm sinh}^2 \chi \Phi^0_\beta (\chi)
& K < 0 \\
\frac{{\rm d}}{{\rm d} \chi} \left[ \frac{\chi^2}{\beta} \Phi^1_\beta (\chi)
\right] = \chi^2 \Phi^0_\beta (\chi) & K = 0 \\
\frac{{\rm d}}{{\rm d} \chi} \left[ \frac{{\rm sin}^2 \chi}{\sqrt{\beta^2 -
1}} \Phi^1_\beta (\chi) \right] = {\rm sin}^2 \chi \phi^0_\beta
(\chi) & K > 0 
\end{array} \; . 
\label{relat-1}
\end{eqnarray}

The vector eigenfunctions satisfy (\ref{eqn-Helm}) for $|m| = 1$,
subject to the divergenceless condition $D^i Q^{(\pm 1)}_i = 0$,
 where the spectrum
is defined by $\beta^2 = (k^2 + 2 K)/|K|$. The radial component $Q^{(\pm1)}_1$ is given by
\begin{eqnarray}
Q^{(1)}_1 = \frac{1}{\beta} \left[ l (l +1) \right]^{1/2} \Phi^l_\beta
Y_{lm} \left\{ 
\begin{array}{ll} \frac{1}{\sinh \chi} & K < 0 \\
 \frac{1}{\chi} & K = 0 \\ \frac{1}{\sin \chi} & K > 0 \end{array}
\right.  \label{q-vect}
\end{eqnarray}
and similarly for $m = -1$, while the remaining (angular) components may be obtained by
substituting (\ref{q-vect}) into (\ref{eqn-Helm}).  The vector
eigenfunctions satisfy normalisation conditions obtained by replacing
$\zeta^*_{\beta^\prime l^\prime m^\prime} \zeta_{\beta l m}$ with
$\zeta^{* (\pm 1)}_{i \;\; \beta^\prime l^\prime m^\prime} \zeta^{i
\;\; (\pm
1)}_{\beta l m}$ in (\ref{normcond}), with $Q^{(\pm 1)}_{i} (\beta,
l, m) = \zeta^{(\pm 1)}_{i \;\; \beta l m}$. 

The tensor eigenfunctions satisfy (\ref{eqn-Helm}) for $|m| = 2$, 
 subject to the transverse-traceless  ($D^i Q^{(2)}_{ij} = 0 =
  Q^{(2) \;\; i}_{i}$)  and symmetry ($Q^{(2)}_{ij} =
Q^{(2)}_{ji}$)  constraints.  The spectrum is given by $\beta^2 = (k^2 + 3K)/|K|$.  The radial--radial
component is given by
\begin{eqnarray}
Q^{(2)}_{11} = \left[ \frac{ (l + 2)(l + 1) l (l - 1)}{2 \beta^2
(\beta^2 - K)} \right]^{1/2} \Phi^l_\beta Y_{lm} \left\{
\begin{array}{ll}
\frac{1}{\sinh^2 \chi} & K < 0 \\
\frac{1}{\chi^2} & K = 0 \\
\frac{1}{\sin^2 \chi} & K > 0
\end{array} \right. \label{qten}
\end{eqnarray}
and similarly for $m = -2$.  The remaining components may be obtained
by substituting (\ref{qten}) in (\ref{eqn-Helm}) and the constraints.
  Normalisation
has again been achieved by a replacement of the integrand in
(\ref{normcond}) by $\zeta^{* (\pm 2)}_{ij \;\; \beta^\prime l^\prime
m^\prime} \zeta^{ij \;\; (\pm 2)}_{\beta l m}$. 

Using this decomposition we may decompose arbitrary  scalar, vector and tensor
perturbations. For example, the pseudo-tensor $\tau^\mu_{\;\; \nu}$
defined in (\ref{tau-def1}) may
be written as:
\begin{eqnarray}
\tau^0_{\;\; 0} ({\bf x}, \tau) &=& \int {\rm d} \mu (\beta) \tau_S  Q^{(0)} \; , \nonumber \\
\tau^0_{\;\; i} ({\bf x}, \tau) &=&  \int {\rm d} \mu (\beta) \left[ \tau_{IV}
 Q^{(0)}_i + 
 \tau^{(1)}_{V} Q^{(1)}_i + \tau^{(-1)}_{V} Q^{(-1)}_i \right] \; ,
\label{met-decomp} \\
\tau^i_{\;\; j} ({\bf x}, \tau) &=& \int {\rm d} \mu (\beta) \left[ 2
\left( \tau_L \gamma^i_{\;\; j} Q^{(0)} +
\tau_T Q^{(0) \; i}_{\;\;\;\;\;\;\;\; j} \right) + \tau_{IT}^{(1)} Q^{(1) \; i}_{\;\;\;\;\;\;\;\; j}
+ \tau^{(-1)}_{IT} Q^{(-1) \; i}_{\;\;\;\;\;\;\;\;\;\; j}  \nonumber \right. \\
& & \;\; + \left. \tau_G^{(2)} Q^{(2) \; i}_{\;\;\;\;\;\;\;\; j} +
\tau_G^{(-2)} Q^{(-2) \; i}_{\;\;\;\;\;\;\;\;\;\; j} \right] \; ,
\nonumber 
\end{eqnarray}
where the coefficients on the right hand side have suppressed
dependence on both $\beta$ and $\tau$. Their explicit form is given in
ref. \cite{Amery3}.  
Here $\tau_L$ and $\tau_T$ represent two `longitudinal' and `transverse'
scalar degrees of freedom, associated with the auxillary tensors
$\gamma^i_{\;\; j} Q^{(0)}$ and $Q^{(0) \; i}_{\;\;\;\;\;\;\;\; j} = k^{-2}
Q^{(0) \; |i}_{\;\;\;\;\;\;\;\;  |j} + \frac{1}{3} \gamma^i_{\;\; j} Q^{(0)}$
respectively; $\tau_{IV}$ is the vector
mode `induced' by the auxillary vector $Q^{(0)}_i = -
k^{-1}Q^{(0)}_{|i}$; and  $\tau_{IT}^{(\pm1)}$ are the tensor modes
`induced' by $Q^{(\pm1)}_{ij} =- (2k)^{-1} \left[ Q_{i|j}^{(\pm 1)} + Q_{j|i}^{(\pm 1)}
\right]$.   
As well as the transform over the 
`radial' coordinate $\beta$, there is an implicit sum over indices
$\ell m$ which label the spherical harmonics encoding the angular
dependence.

\section{``Addition'' results for $\Phi^{\ell}_\beta
(\chi)$} \label{App-III}

Here we generalise the result (\ref{j0-genbreak}) to the $K \not= 0$ and 
$l \not= 0$ cases, and prove that the radial 
eigenfunctions at ${\bf u}$ and ${\bf v}$ are related to those at ${\bf u + v}$ 
by
\begin{eqnarray}
\frac{\Phi^n_\beta (\chi_{\bf u + v})}{(\sinh \chi_{\bf u + v})^n } =
\Omega_\nu (\chi_{\bf u + v}) = \sum^\infty_{m = 0} b_m
\frac{\Phi^{n + m}_\beta (\chi_{\bf u})}{(\sinh \chi_{\bf u})^n}
\frac{\Phi^{n + m}_\beta (\chi_{\bf v})}{(\sinh \chi_{\bf v})^n}
C^\nu_m (\cos \sigma) \; , 
\label{AII-8}
\end{eqnarray}
where $b_m$ is independent of $\chi_{\bf u}$, $\chi_{\bf v}$ and
$\sigma$, and may be found by comparing coefficients. For 
${\bf u}= -{\bf x}$ and ${\bf v} = {\bf x}+{\bf y}$ and for 
the $l = 0$
case, we may appeal to the
common normalisation of our radial eigenfunctions (for the $K = 0$ and
$K \not= 0$ universes) to obtain (\ref{Phi0-genbreak}).

 Foch \cite{foch} proved this result for the special case $l = 0$, and
for a closed universe only. Here we follow the method of
Gegenbauer \cite{gegen}  to obtain a more
general result for the $l$-th radial eigenmode of an open
universe. The proof for the closed universe is similar. 

For curved geometries,  the radial
coordinates at ${\bf u}$ and ${\bf v}$ are related to those 
at ${\bf u} + {\bf v}$ via
\begin{eqnarray}
{\rm cos}_K \chi_{{\bf u} + {\bf v}} = {\rm cos}_K \chi_{\bf u} {\rm
cos}_K \chi_{\bf v} - \frac{K}{|K|} {\rm sin}_K \chi_{\bf u} {\rm sin}_K \chi_{\bf v} {\cos} \sigma
\; , \label{eep-1.1}
\end{eqnarray}
where we use $\sin_K\chi$ and $\cos_K \chi$ as previously defined,
but only for the $K\not= 0$ cases.  
Here,  $\sigma$ is the angle between ${\bf u}$ and ${\bf v}$ (in the
hyperplane defined by these vectors), and where
we may (without loss of generality) choose ${\bf u}$ pointing towards
the north pole such that $\sigma = \theta$. For the open (closed)
universes, equation (\ref{eep-1.1}) 
 clearly reduces to the flat space result $r_{{\bf u} + {\bf v}}^2
= r_{\bf u}^2 + r_{\bf v}^2 + 2 r_{\bf u} r_{\bf v} {\rm cos} \sigma$,
in the local limit and for $|K| \longrightarrow 0$. 

The $n$-th radial eigenfunction satisfies the equation
\begin{eqnarray}
\frac{\partial^2 \Phi^n_\beta (\chi)}{\partial \chi^2} + 2 \coth \chi
\frac{\partial \Phi^n_\beta (\chi)}{\partial \chi} + \left( k^2 -
\frac{n (n + 1)}{\sinh^2 \chi} \right) \Phi^n_\beta (\chi) = 0 \; . 
\label{AII-1}
\end{eqnarray}
From the $\Phi^n_\beta$  we may construct the quantity
\begin{eqnarray}
\Omega_\nu \equiv \frac{\Psi_\nu}{(\sinh \chi )^\nu} = \frac{\Phi^n_\beta
(\chi)}{(\sinh \chi)^n} \; , 
\label{AII-2}
\end{eqnarray}
where $\nu = n + 1/2$ and $\Psi_\nu$, defined by $\Psi_\nu \equiv \sqrt{\sinh \chi} \Phi^n_\beta
(\chi)$, expresses a relationship similar to that between the spherical
(half-integer) and ordinary (integer) Bessel functions. Then,
$\Omega_\nu$ satisfies the equation 
\begin{eqnarray}
\frac{\partial^2 \Omega_\nu}{\partial \chi^2} + (2 \nu + 1) \coth \chi
\frac{\partial \Omega_\nu}{\partial \chi} + [ k^2 + (\nu - 1/2) (\nu +
3/2) ] \Omega_\nu = 0 \; . 
\label{AII-3}
\end{eqnarray}
Next, we use the addition rule for a hyper-sphere (\ref{eep-1.1}) 
 to write
(\ref{AII-3}) in terms of derivatives with respect to $\chi_{\bf u}$
and $\sigma$, assuming without loss of generality that $\chi_{\bf v}$ is
constant. Hence, we obtain the equation
\begin{eqnarray}
\frac{1}{\sinh^2 \chi_{\bf u}} \left[ \frac{\partial^2 }{\partial
\sigma^2} + 2 \nu \cot \sigma \frac{\partial }{\partial \sigma}
\right] \Omega_\nu &+& \frac{\partial^2 \Omega_\nu}{\partial \chi_{\bf
u}} + (2 \nu + 1) \coth \chi_{\bf u} \frac{\partial
\Omega_\nu}{\partial \chi_{\bf u}} \nonumber \\
&+& \left[ k^2 + (\nu - 1/2) (\nu +
3/2) \right] \Omega_\nu = 0  \; . 
\label{AII-5}
\end{eqnarray}

We now assume that $\Omega_\nu$ can be expanded as 
 $\Omega_\nu = \sum^\infty_{m = n} B_m C^\nu_m (\cos \sigma)$ 
where $B_m$ is independent of $\sigma$, and the $C^\nu_m (\cos \sigma)$ is
a (Gegenbauer) polynomial of degree $m$ in $\cos \sigma$. Formally,
it may be shown \cite{gegen} that $\left\{ \frac{\partial^2}{\partial \sigma^2} + 2 \nu
\cot \sigma \frac{\partial }{\partial \sigma} \right\} C^\nu_m (\cos
\sigma)$ is a constant multiple of $C^\nu_m (\cos \sigma)$, so that
$C^\nu_m (\cos \sigma)$ may be taken to be the coefficient of $a^m$ in
the series expansion of $( 1 - 2 a \cos \sigma + a^2)^{-\nu}$.  

From the properties of these 
Gegenbauer functions $C^\nu_m (\cos \sigma)$, \cite{abram}, we thus arrive at the
following equation for $B_m$, as a function of $\chi_{\bf u}$:
\begin{eqnarray}
\frac{\partial^2 B_m}{\partial \chi_{\bf u}^2} + (2 \nu + 1) \coth
\chi_{\bf u} \frac{\partial B_m}{\partial \chi_{\bf u}} + [ k^2 + (\nu
- 1/2) (\nu + 3/2) ] B_m - \frac{m (2 \nu + m)}{\sinh^2 \chi_{\bf u}}
B_m = 0 \; , \nonumber \\
\label{AII-7}
\end{eqnarray}
which has solutions 
$B_m \propto \frac{\Phi^{n + m}_\beta (\chi_{\bf u})}{(\sinh \chi_{\bf
u})^n}$, 
and so, appealing to the symmetry between $\chi_{\bf u}$ and $\chi_{\bf v}$,
we have (\ref{AII-8}).

\end{document}